\documentclass[namedreferences]{solarphysics}

\usepackage[hyperref,optionalrh]{spr-sola-addons}
\usepackage{graphicx}        
\usepackage{color}           
\usepackage{breakurl}        

\newcommand{\ux}{$U_x$}
\newcommand{\uy}{$U_y$}
\newcommand{\rdfitc}{{\tt rdfitc}}
\newcommand{\rdfitf}{{\tt rdfitf}}

\newcommand{\aap}{    {\it Astron. Astrophys.}}

\newcommand{\apj}{    {\it Astrophys. J.}}

\newcommand{\solphys}{{\it Solar Phys.}}

\chardef\us=`\_
\newcommand{\rsun}{R$_\odot$}
\begin{document}

\begin{article}
\begin{opening}

\title{HMI Ring Diagram Analysis:\\ Effects of tracking, noise and resolution}

\author[addressref={aff1},corref,email={sarbani.basu@yale.edu}]{\inits{S.B.}\fnm{Sarbani}~\lnm{Basu}\orcid{0000-0002-6163-3472}}
\author[addressref=aff2,email={rbogart@stanford.edu}]{\inits{R.S}\fnm{Richard S. }~\lnm{Bogart}\orcid{0000-0002-0910-459X}}

\address[id=aff1]{Department of Astronomy, Yale University, PO Box 208181, New Haven, CT 06520-8101, USA}
\address[id=aff2]{W. W. Hansen Experimental Physics Laboratory, Stanford University, Stanford, CA, 94305-4085, USA}

\runningauthor{Basu \& Bogart}
\runningtitle{Ring diagram analysis}

\begin{abstract}
Ring diagram analysis is a standard local helioseismic technique. Data from the Helioseismic and Magnetic Imager (HMI) on the Solar Dynamics Observatory (SDO) are routinely used for Ring-diagram analysis, and fits to the power spectra as well as inversion results are standard data products. In this paper we examine the effects of different tracking rates, noise, and resolution on ring-diagram results. Most of the analysis is  for $15^\circ$ tiles, but we also examine the effects of different tile sizes. The largest effect we find is that of resolution.  Doppler noise has very little effect on the results, except perhaps at the deepest regions for which the tiles can give reliable results; variations in the tracking rate have a similar effect. 

\end{abstract}
\end{opening}

\section{Introduction}
     \label{sec:introduction} 

Ring-diagram analysis has been used to study large-scale flows in the outer regions of the Sun.
This technique involves using high-degree
$p$-modes from  3-dimensional (3D) power spectra that are obtained from
parts of the solar surface which are tracked with a known rate \citep{frankhill}.
The 3D spectra are fit with a model to obtain mode frequencies, 
as well as two velocity parameters $U_x$ and $U_y$, representing horizontal displacement of the centers of the rings in the frequency plane corresponding to different modes; this displacement is caused by horizontal flows. These parameters are inverted for the depth dependence
of the transverse velocity of the observed field (relative to the tracking rate); conventionally, $U_x$ is chosen to be the zonal direction and $U_y$ the meridional direction. These velocity parameters have been used extensively to study solar near-surface flows and their evolution \citep[etc.]{haber2000, irene2006, basu2010, jain2012, persistent, lekshmi2018, hanson2020, komm2021}.
While most ring-diagram results are for solar flows, there has also been a 
limited amount of work on the structure of the near-surface layers of the Sun.
\citep[e.g.][]{basu2004,basu2007,bogart2008,baldner2013}.

The Helioseismic and Magnetic Imager \citep[HMI;][]{HMI} on the Solar Dynamics Observatory (SDO) is the primary source of high-cadence, high-spatial resolution data suitable for ring-diagram analyses. Prior to this, the Michelson Doppler Imager \citep[MDI;][]{mdi} on the Solar and Heliospheric Observatory was the source of space-based data, but at a lower spatial resolution.
The ground-based GONG project \citep{gong} also produces ring-diagram data, but at a lower spatial resolution than HMI.
The HMI project routinely produces tracked cubes of different spatial sizes and durations, power spectra, fits to the power spectra, as well as flow inversions. Documentation  on the pipeline analysis modules and associated data products can be found on the web pages of the HMI Ring Diagrams Team.\footnote{http://hmi.stanford.edu/teams/rings/} We use the HMI pipeline for tracking and fitting the power spectra used in our analysis.

The rest of the paper is organized as follows. In Section~\ref{sec:data} we describe the data analysis. Section~\ref{sec:15deg} discusses the analysis of and results obtained with standard $15^\circ$ tiles. This section describes how the tracking rate, noise and pixel resolution affect flow inversion results. Section~\ref{sec:30deg} describes the effect of noise on $30^\circ$ tiles. In Section~\ref{sec:5deg} we show that it is possible to invert data obtained by fitting $5^\circ$ tiles --- inversion results from these tiles are not an HMI data product; the pipeline only provides frequencies and the fit parameters.  In this section, we also examine the effects of noise on flows obtained from $5^\circ$ data. Finally, in Section~\ref{sec:disc} we discuss the implications of the results.

\section{Data and Data processing}
\label{sec:data}

The HMI ring-diagram pipeline and the resulting data products are described in \citet{ringpipe1,ringpipe2}. Briefly, the photosphere is tiled with sets of mapped overlapping circular analysis regions in the Dopplergrams centered at fixed Carrington coordinates, of diameters $5^\circ$, $15^\circ$, and $30^\circ$ respectively, each tracked for about the time it takes to rotate through its diameter. The spatial-temporal power spectra of the resultant data cubes are fit with two different models for the peaks corresponding to acoustic-mode ridges. One model, \rdfitc, is based on that of \citet{rdfitc} and fits asymmetric
profiles to the peaks. The anisotropies in the distribution of power around a ring are fitted explicitly. The fits are done with frequency as the independent variable. The process is slow, but results in a large number of fitted parameters. Also, because the frequency resolution associated with the comparatively long tracking times is high, the fits somewhat oversample the data. The second is the model \rdfitf\ based on \citet{rdfitf}\, where a symmetric profile is fit to the peaks. In this case, the anisotropy around a ring is removed by remapping the horizontal wavelength from a Cartesian $k_x,k_y$ grid to a $k_r, k_\theta$ polar grid, then averaging the rings in frequency, determining the Fourier components and dividing out the frequency independent part of the signal. To speed up the code, the remapped rings in $\theta$ are Fourier transformed, and only the low-order coefficients are retained. The angular transform is inverted back to $\theta$ coordinates and the data are sub-sampled to a coarser grid in azimuth. This makes the process fast, but also means that the number of fitted modes can be very small. Additionally, as we show below, the uncertainties in the flow parameters are overestimated.

The fitted \ux\ and \uy\ parameters are inverted using the Optimally Localized Averages (OLA) method. In some cases, as a comparison, we also use the Regularized Least Squares (RLS) method. OLA and RLS inversions are complementary in nature \citep{sekii1997}, and inversions can be trusted if both inversion techniques return the same results. RLS aims to find values of \ux\ and \uy\ that give the best fit to the data (i.e., give the smallest residuals) while keeping uncertainties small; the aim of OLA is not to fit the data at all, but to find linear combinations of the velocity in such a way that the corresponding combination of kernels, the \textit{averaging} or \textit{resolution} kernel, provides a localized average of the underlying flow, again while keeping uncertainties small. The techniques, as applied to ring-diagram data, are described in \citet{ringsinv}.

Ring-diagram analysis at the scale of the $5^\circ$ tiles is possible with the comparatively high spatial resolution of HMI. Similar analyses based on data from GONG and MDI were generally restricted to tiles of $15^\circ$ diameter. Also, because of the volume of data involved, the HMI $30^\circ$ tiles are mapped at only half the spatial resolution of the others. Consequently, most of our work is focused on the results obtained with $15^\circ$ tiles.
For these, we chose to examine data from two periods, each involving five successive time samples, centered around times of $B_0 \sim 0$, one during solar minimum and one during solar maximum. {\bf (We also included five time samples from a time of low solar activity near the start of the SDO mission when $B_0$ was not so close to 0.)} The length of the tracking intervals for each sample depends upon the tile size. For the $15^\circ$ tiles in the HMI pipeline it is 2304 45-sec images, corresponding to an interval of $28^{h}48^m$. (For the MDI and GONG pipelines the corresponding length is 1664 1-minute images for an interval of $27^{h}44^m$.) For each of the {fifteen} time targets for the $15^\circ$ study, we analyzed 11 regions: five on the equator at central meridian distances of $0^\circ$.  $\pm15^\circ$, and $\pm30^\circ$; and six on the central meridian at latitudes $\pm30^\circ$, $\pm60^\circ$, and $\pm67.5^\circ$. These target times and characteristics of the associated regions, are described in Table \ref{tbl:targets}.

\begin{table}
\begin{tabular}{cc cc cc rr cc}
Target && MidTime && $B_0$ && MAI && Location \\
{2097:315} && 2010.05.23 09:33:22.5 && $-$1.715 && 1.808 && 315.0+30.0 \\
{2097:300} && 2010.05.24 12:45:22.5 && $-$1.582 && 1.894 && 300.0+30.0 \\
{2097:285} && 2010.05.25 15:58:07.5 && $-$1.448 && 1.197 && 285.0+00.0 \\
{2097:270} && 2010.05.26 19:10:52.5 && $-$1.314 && 1.239 && 300.0+00.0 \\
{2097:255} && 2010.05.27 22:22:52.5 && $-$1.179 && 2.465 && 255.0$-$30.0 \\
2157:030 && 2014.12.06 07:57:22.5 && +0.233 && 21.811 && 000.0+00.0 \\
2157:015 && 2014.12.07 11:16:52.5 && +0.087 && 27.394 && 000.0+00.0 \\
2158:360 && 2014.12.08 14:35:37.5 && $-$0.059 && 27.365 && 000.0+00.0 \\
2158:345 && 2014.12.09 17:55:07.5 && $-$0.205 && 25.753 && 000.0+00.0 \\
2158:330 && 2014.12.10 21:13:52.5 && $-$0.350 && 26.904 && 000.0+00.0\\
2224:060 && 2019.12.05 11:55:52.5 && +0.375 && 1.376 && 060.0+00.0 \\
2224:045 && 2019.12.06 15:14:37.5 && +0.229 && 1.424 && 060.0+00.0 \\
2224:030 && 2019.12.07 18:34:07.5 && +0.084 && 1.508 && 015.0+00.0 \\
2224:015 && 2019.12.08 21:52:52.5 && $-$0.062 && 1.707 && 015.0$-$30.0 \\
2225:360 && 2019.12.10 01:12:22.5 && $-$0.208 && 1.499 && 000.0$-$30.0 
\end{tabular}
\caption{
Description of target times selected for analysis of $15^\circ$ tiles. For each target in Carrington Time (Carrington Rotation and longitude of central meridian), the midpoint of the tracking interval is given, along with the heliographic latitude (in degrees) of disc center at that time.  The last two columns give the maximum value of the Magnetic Activity Index (MAI) associated with any of the $15^\circ$ tiles sampled at that time and the Carrington coordinates of the associated tile. For the five times during solar maximum, those correspond to a tile that contained Active Region AR 12227.
}
\label{tbl:targets}
\end{table}
For the studies using $30^\circ$ tiles, which are tracked twice as long ($57^{h}36^{m}$) but only half as often, we used the three target Carrington times during solar minimum 2224:060, 2224:030, and 2225:360, with regions on the equator at central meridian distances $0^\circ$, $\pm30^\circ$, and on the central meridian at latitudes $\pm45^\circ$. For the $9^{h}36^{m}$ trackings of $5^\circ$ tiles, we examined the seven Carrington times from 2224:035 (2019.12.07 16:33) -- 2224:005 (2019.12.09 16:12) sampled every 5$^\circ$ in rotation. For each of these times we analyzed regions on the central meridian at latitudes $0^\circ$, $\pm30^\circ$, $\pm45^\circ$, and 
$\pm60^\circ$. Apart from the $15^\circ$ tile containing AR 12227 in 2014 noted in Table \ref{tbl:targets}, all of the regions studied were magnetically quiet as measured by the associated MAI, a measure of the average line-of-sight component of magnetic flux in the region during the tracking interval.

\begin{figure}
\centering
\includegraphics[width=0.9\textwidth]{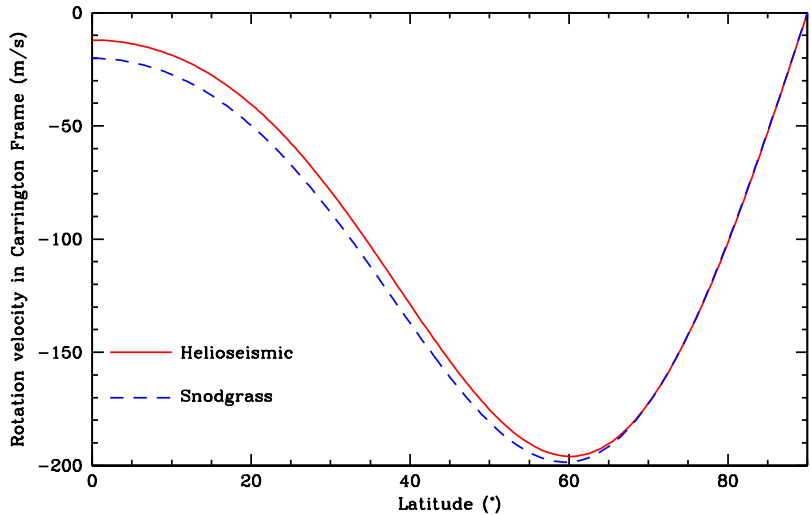}
\caption{
Comparison of a differential rotation model based on HMI $5^\circ$ ring-diagram $f$-mode fits averaged over a year around solar minimum (solid red curve; the``helioseismic'' rate) and the ``Snodgrass'' rate based on photospheric Doppler measurements over about 2 years around solar maximum (dashed blue curve).
The rotation rate for the models as a function of latitude is displayed in the Carrington coordinate system, the $U_x$ value expected in the absence of a local zonal flow when tracking at the Carrington rate. The rotation model is the standard
$\Omega = A_0 + A_2\ \sin^{2}\lambda  + A_4\ \sin^{4}\lambda,$
where $\lambda$ is the heliographic latitude. The coefficients for the Snodgrass rate, expressed in units of $\mu {\rm Rad\ s}^{-1}$, are $A_0 = 2.8364$, $A_2 = -0.3441$, $A_4 = -0.5037$; for the helioseismic rate they are $A_0 = 2.8480$, $A_2 = -0.3160$, $A_4 = -0.5491$.
}
\label{fig:cfdrmods}
\end{figure}
The measured flows inferred from the Doppler displacement of the acoustic modes are of course with respect to the frame of reference, which depends on the tracking rate (and direction!). For the MDI and GONG pipelines, the aim was to keep the displacements small by tracking at a rate corresponding to the expected differential rotation appropriate to the latitude of the tile center. For this purpose, a differential rotation model inferred from surface Doppler measurements \citep{snodgrass} was adopted. In the HMI pipeline, the tiles are tracked at a uniform rate consistent with their Carrington coordinates. To the extent that measured flows, after correction for the frame of reference, differ in the two cases, these differences may be attributable to the effect of tracking at the different rates. On the other hand, they might simply be due to the sensitivity, especially of the mode-fitting procedures, to different instantiations of similar populations. To examine this issue, we have not only compared results from the same HMI data tracked at both the Carrington rate and the so-called ``Snodgrass'' rate, but also from data tracked at a slightly different rate. To that end, we have used means of the $f$-mode fits of $5^\circ$ tiles over a full year around the time of solar minimum to estimate coefficients for a similar near-surface differential rotation model. This model, which might be termed a ``helioseismic'' surface differential rotation, is about 10 m/s faster at the equator than the photospheric rate, which is to be expected due to its slightly deeper sensitivity. It also however has a slightly steeper latitude dependence --- the two models cross around latitude 70$^\circ$. This is likely attributable to the fact that our model is based on data taken during solar minimum, while the observations on which the ``Snodgrass'' model is based were made around a time of solar maximum. If we use HMI $f$-mode data from a year around solar maximum, the values for the derived $A_2$ and $A_4$ coefficients (see Fig.~\ref{fig:cfdrmods}) agree very closely with those in the ``Snodgrass'' model. { The derived $A_0$ coefficient representing the equatorial rotation rate is effectively constant over the cycle, and it accounts for most of the difference between the ``Snodgrass'' and our helioseismic model.}

In order to investigate the sensitivity of flows inferred from ring-diagram analysis to noise in the measured data, we have taken the tracked data cubes used in the original HMI pipeline analysis for the selected target regions, and added normally-distributed random noise centered at 0 to the value at each voxel. The per pixel noise in the original Dopplergrams ranges from about 15 m/s at disc center to about 50 m/s near the limb \citep{hmical}. Since the mapping involved in the tracking involves a cubic-convolution interpolation and is close in scale to the observational pixel resolution (except for $30^\circ$ tiles), the per-voxel noise in the tracked cubes can be assumed to be somewhat less or comparable over most of the disc, on the order of 10 m/s. (By way of comparison, the total RMS per image pixel, including both acoustic mode signal and solar ``noise'' such as granulation and super-granulation, is about 360 m/s near disc center and 500 m/s at $\mu \sim 0.5$.) To these remapped and tracked data we have added noise with RMS values of 1, 3, 10, 30, and 100 m/s respectively before repeating the rest of the pipeline analysis.


We have experimented with two simple methods of reducing the spatial resolution of HMI Doppler data to investigate the effects upon the inferences from the ring-diagram analyses: {\it a)} interpolating to the mapped tracked cubes at lower resolution; and {\it b)} binning the original Dopplergram pixels prior to mapping and tracking. Method {\it a,} which effectively undersamples the original data near disc center, is in fact used in the HMI pipeline for the $30^\circ$ tiles, which are mapped at a resolution of $0.08^\circ$ heliographic per pixel rather than $0.04^\circ$ per pixel, which closely matches the spatial sampling of the HMI images at disc center. Because method {\it b} more closely approximates the data expected from an instrument of lower resolution, we only present those findings here. The fact that a 50\% reduction in resolution with that method has minimal effects (see Section~\ref{sec:15deg}) encourages confidence in the HMI pipeline results for the $30^\circ$ tiles.

\begin{figure}
\centering
\includegraphics[width=0.9\textwidth]{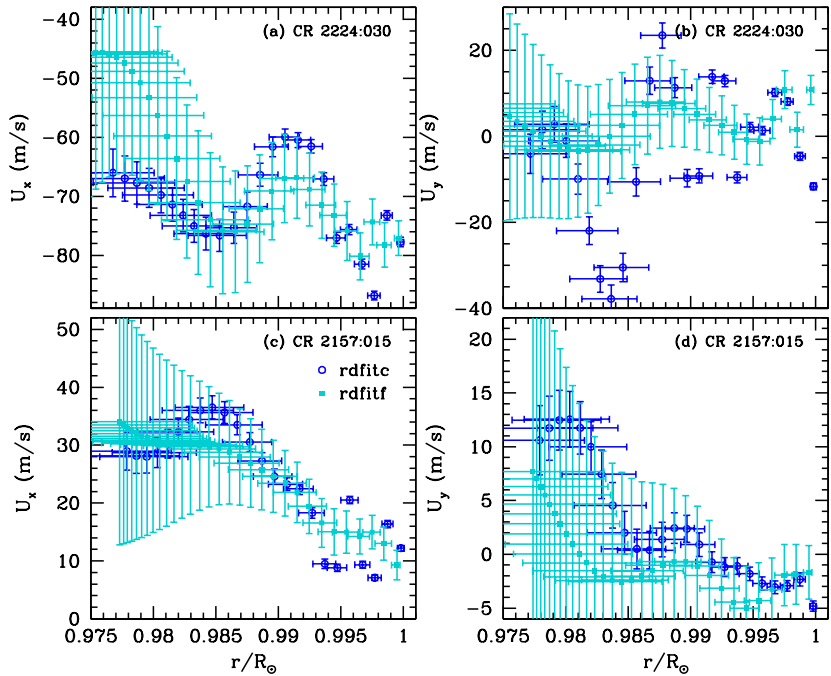}
\caption{A comparison of flows obtained by OLA inversions of \rdfitc\ 
(dark blue) and {\rdfitf} (blue-green) fits
for  $15^\circ$ { tiles at two locations. The top row is for a tile at $30^\circ$N on the central meridian during CR 2224, CM longitude $30^\circ$, the bottom row is for a tile} at { the} disk-center during CR 2157, CM longitude $15^\circ$. The points are plotted at the center of gravity of the averaging kernels. The vertical error-bars show $1\sigma$ uncertainty, while the horizontal error-bars are a measure of the resolution of the inversions, and is the distance between the 25th and 75th quantile of the averaging kernels. Note that uncertainties in the \rdfitf\ results are much larger than those in the \rdfitc\ results.  } 
\label{fig:invcomp}
\end{figure}
\begin{figure}
\centering
\includegraphics[width=0.8\textwidth]{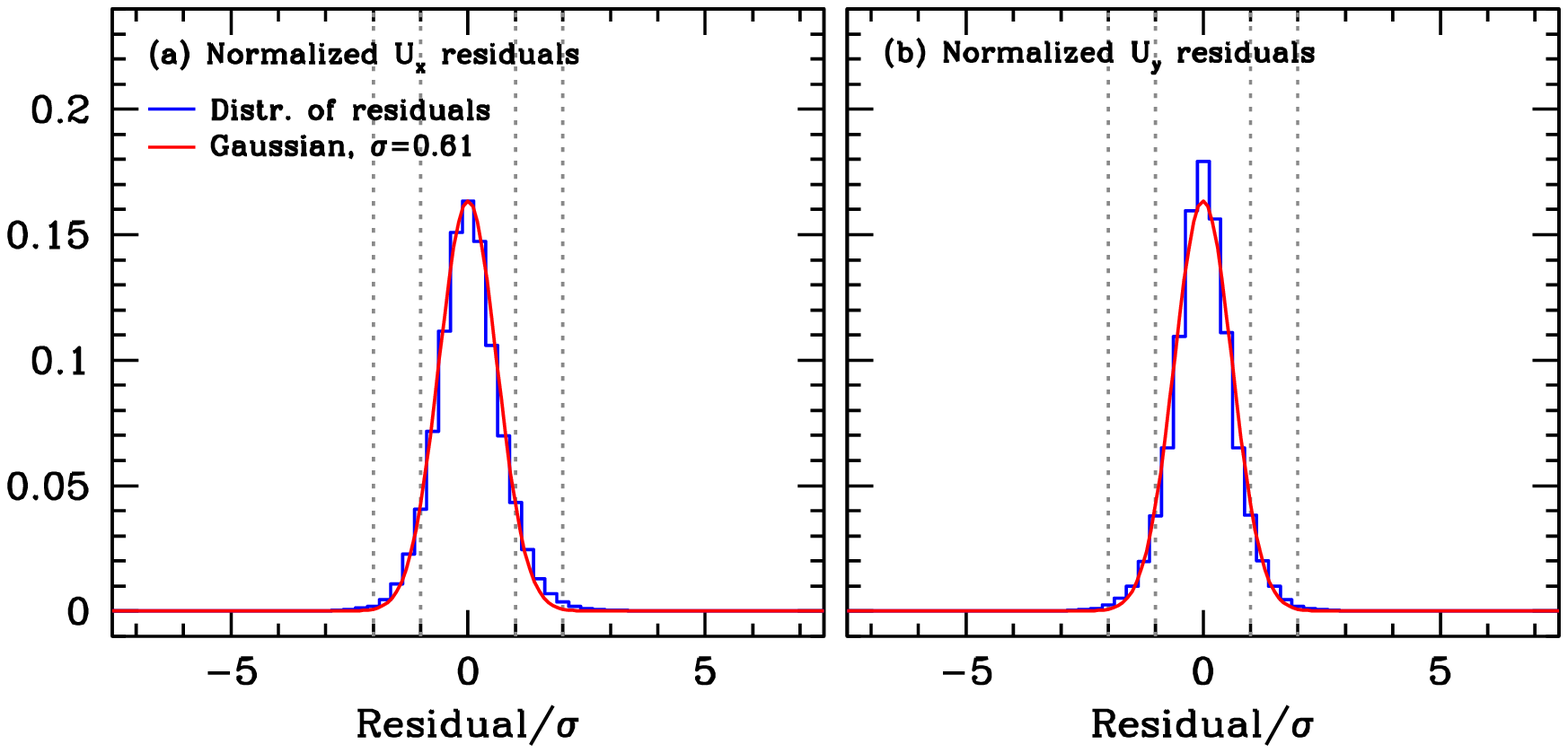}
\includegraphics[width=0.8\textwidth]{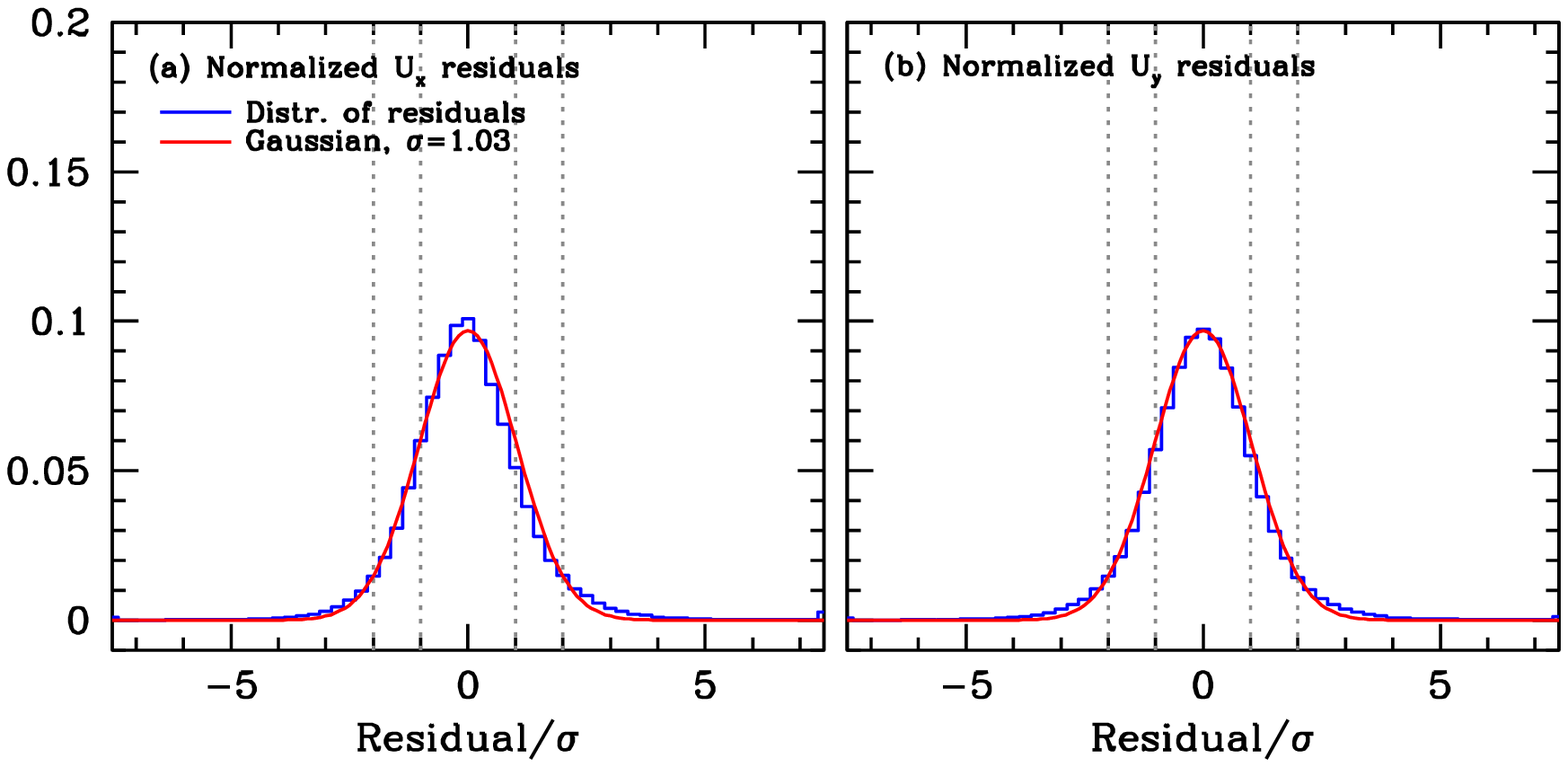}
\caption{{ Top Panel:}The distribution of residuals normalized by uncertainty for RLS inversions of {\rdfitf} parameters of all $15^\circ$ regions are shown as the blue histograms. The red curve shows a Gaussian
with $\sigma=0.61$ fitted to the histogram. The fit to the distribution with $\sigma < 1$ implies that the uncertainties are overestimated, on average, by a factor of 1/0.61.
{ Bottom Panel:} Same as the upper panel, but for {\rdfitc} results. The red curve shows a Gaussian
with $\sigma=1.03$ fitted to the histogram. Thus, on average, \rdfitc\ errors have not been  underestimated. In all panels, vertical dotted lines mark $\pm 1\sigma$ and $\pm 2\sigma$. 
} 
\label{fig:fitf_res}
\end{figure}
\begin{figure}
\centering
\includegraphics[width=0.9\textwidth]{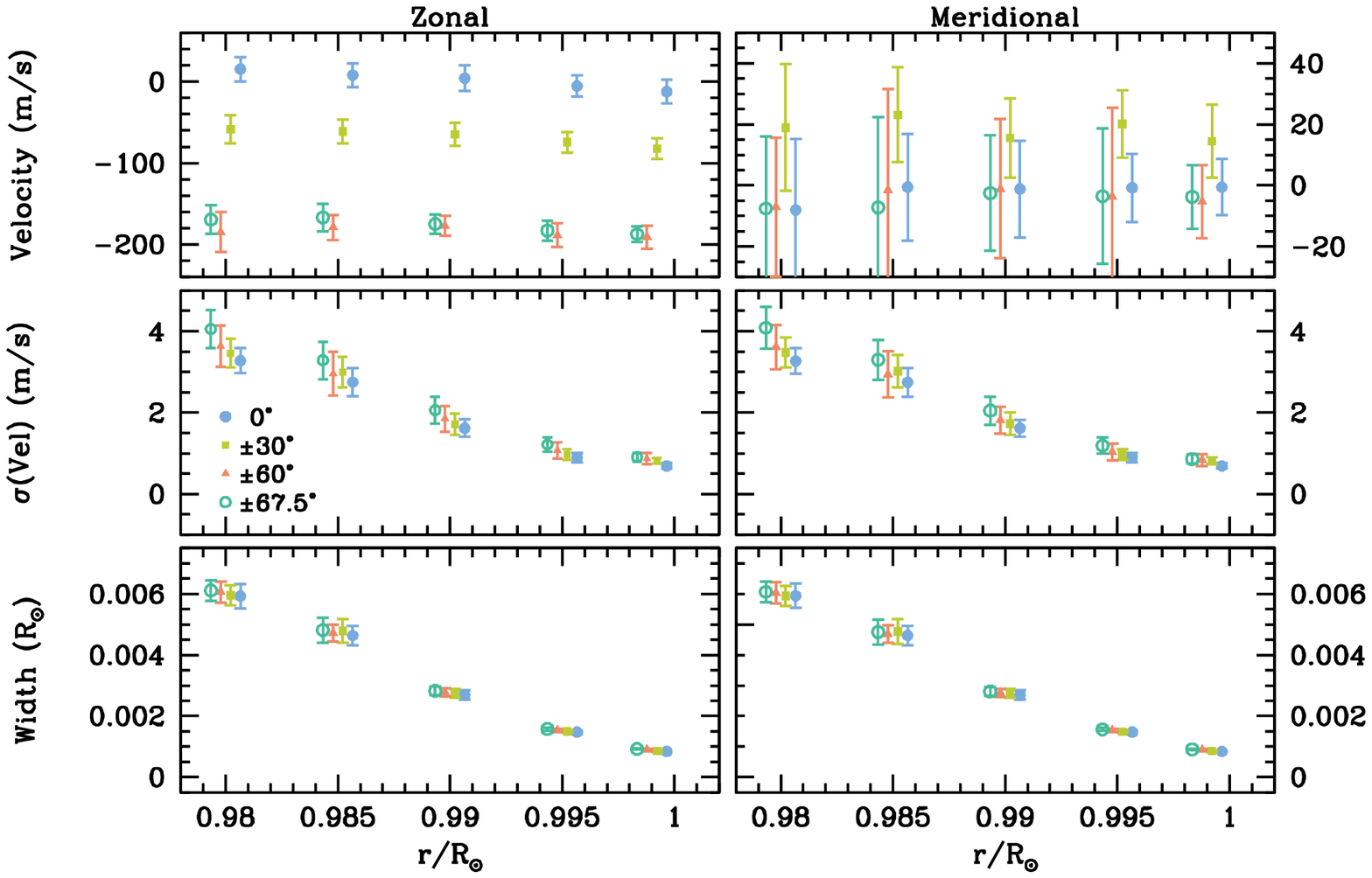}
\caption{The average results obtained by inverting zonal (left column) and meridional (right column) flow parameters; only results along the central meridian are shown. The rows, from top to bottom, show the inverted velocities, uncertainties in the velocities, and the width of the averaging kernel. The vertical error bars in each panel show the $1\sigma$ scatter in the results. The meridional flow velocities are the north-south antisymmetric average of the flows. We show results for inversions centered at 0.98, 0.985, 0.99, 0.995 and 0.999 R$_\odot$. The results for different latitudes are displaced slightly from the true radius for the sake of clarity.
} 
\label{fig:fid}
\end{figure}

\section{Results for 15-degree tiles}
\label{sec:15deg}

In Fig.~\ref{fig:invcomp} we show inversion results for { two} $15^\circ$ { tiles} obtained by inverting modes obtained with both \rdfitc\ and \rdfitf\ fits. The uncertainties obtained with \rdfitf\ are much larger than those with \rdfitc, raising the question whether \rdfitf\ uncertainties are overestimated or whether \rdfitc\ errors are underestimated. To examine this, we look at the distribution of the residuals obtained with RLS inversions of all { 165} $15^\circ$ tiles. If the residuals normalized by uncertainties have a distribution with $\sigma \approx 1$, the estimates of the uncertainties, on average, are correct; if $\sigma$ is considerably less than 1, the uncertainties are overestimated. The distribution of the residuals are shown in Fig.~\ref{fig:fitf_res}, along with Gaussian fits to the residuals.
It is clear that \rdfitf\ uncertainties are overestimates. Given these results, in the rest of the paper, unless otherwise specified, we will only show inversion results for flow parameters obtained with \rdfitc. In Fig.~\ref{fig:fid} we show the results obtained with the standard $15^\circ$ set, which may be considered to be the fiducial results against which we compare results with changed data properties --- tracking rate, noise and spatial resolution.

\subsection{Effects of the tracking rate}
\label{subsec:track}

\begin{figure}
\centering
\includegraphics[width=0.8\textwidth]{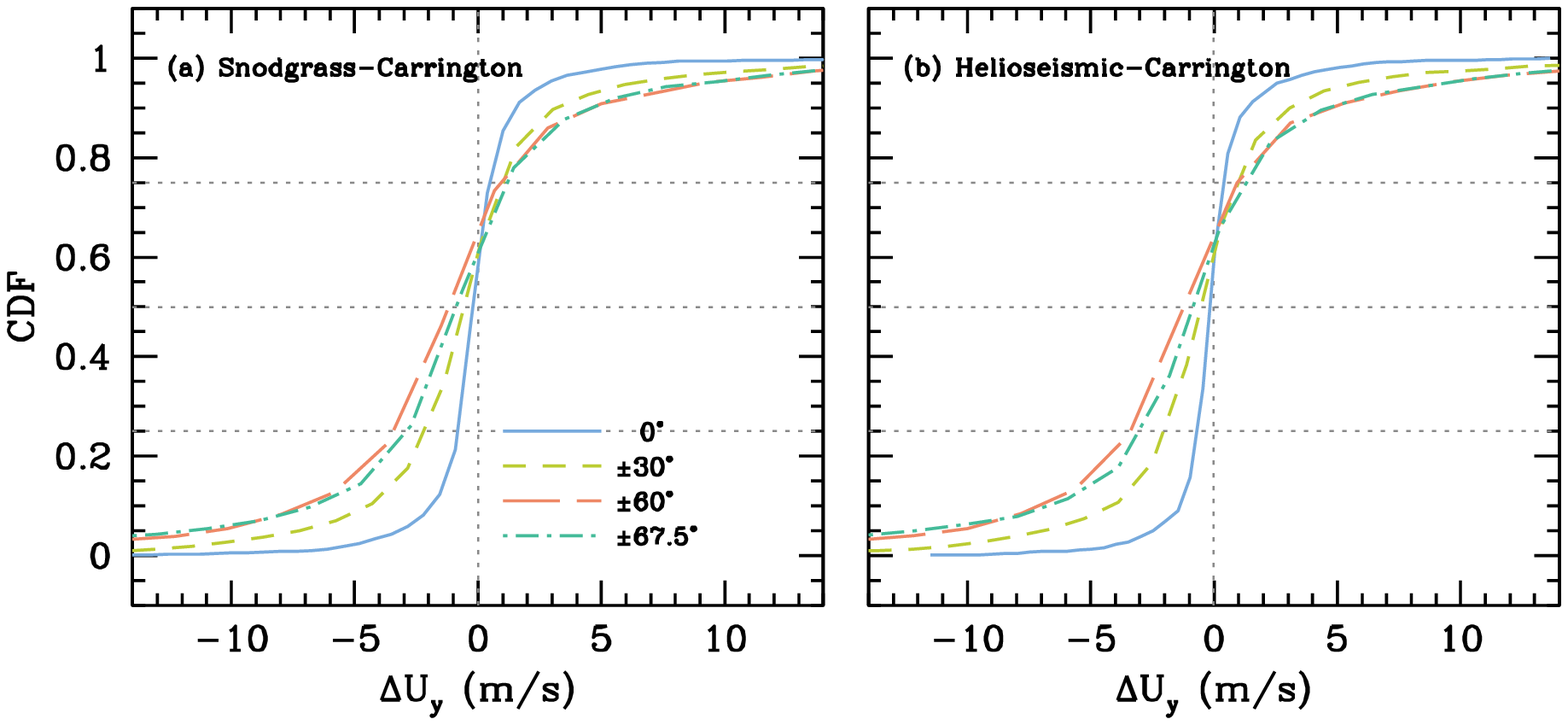}
\centering
\includegraphics[width=0.8\textwidth]{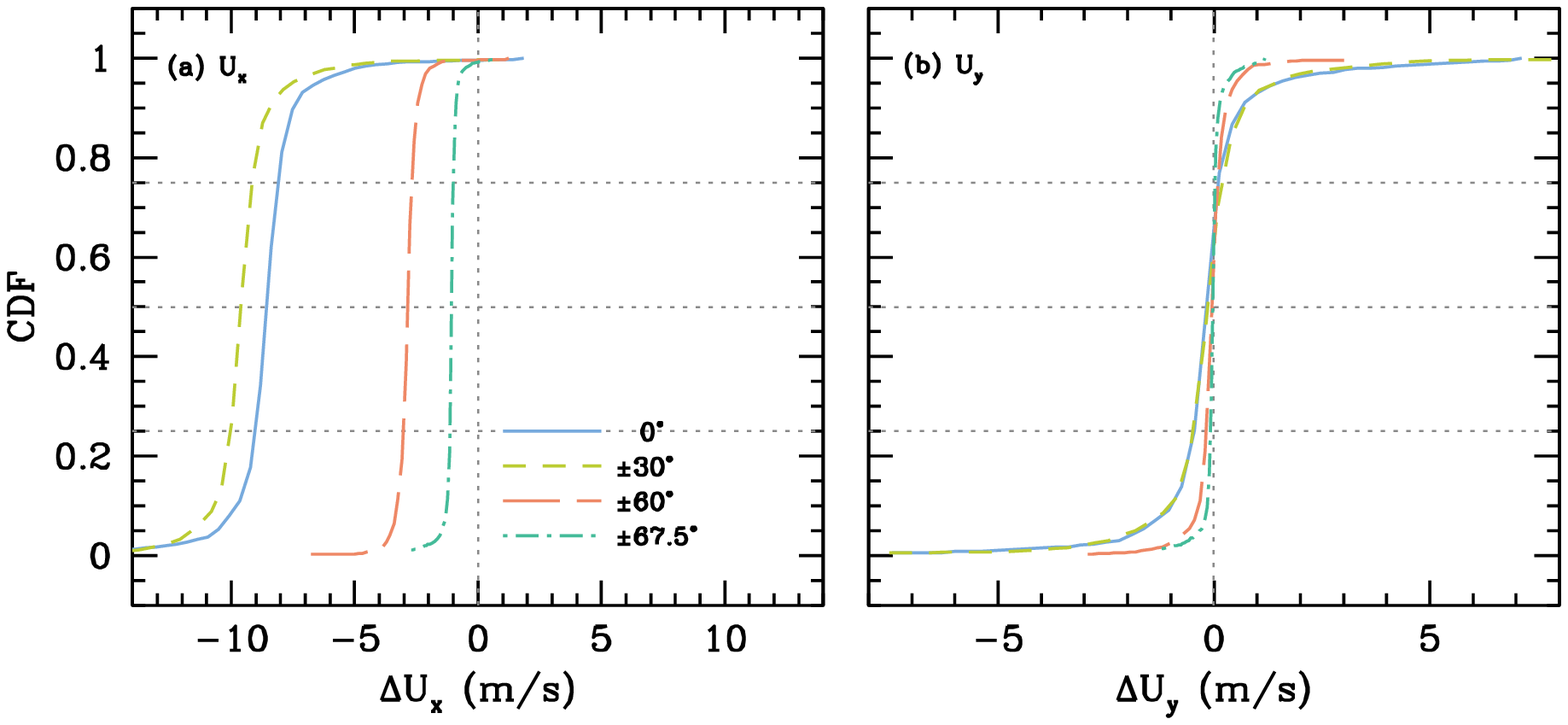}
\caption{The cumulative distribution function (CDF) of the differences between $U_y$ parameters of common modes for regions tracked with different tracking rates. { Top row:} differences with respect to the Carrington rate, in the sense of (other $-$ Carrington), we do not show $U_x$ differences because of the large differences in the tracking rates. { Bottom row:} the CDF of the differences between $U_x$ and $U_y$ parameters of common modes for regions tracked with the Helioseismic and Snodgrass rates. The differences are in the sense of (Helioseismic $-$ Snodgrass).
In all panels, the vertical dotted lines correspond to 0 difference, while the horizontal lines mark the 25th, 50th and 75th quantile. 
} 
\label{fig:cdf_carr}
\end{figure}

As mentioned in Section~\ref{sec:data}, ring diagrams with MDI data were constructed from regions tracked at the Snodgrass rate applicable to their central latitude, while the standard HMI rings are tracked at the Carrington rate. In the top row of Fig.~\ref{fig:cdf_carr} we show the cumulative distribution function (CDF) of the differences in the $U_y$ parameter for the same set of Dopplergrams tracked at Carrington and Snodgrass rates; of course, only the modes that are common between the two sets could be used. We do not show the CDF for the zonal parameter $U_x$, since the difference in tracking rates causes a large shift. Also shown are the differences between Carrington tracking and rings tracked at the helioseismically inferred near-surface rotation rate described above. Note that the differences are smallest at the disc center and become progressively larger at higher latitudes. The differences are also not quite symmetric around zero difference, and the distribution of the differences is skewed somewhat, indicating that the $U_y$ parameter obtained with Snodgrass tracking is lower than that obtained with Carrington tracking. Although this is a systematic difference, the magnitude of the difference is well within $1\sigma$ statistical errors --- only $0.2\sigma$ at a latitude of $60^\circ.$

In the bottom row of Fig.~\ref{fig:cdf_carr} we show the differences between tracking at the Snodgrass and Helioseismic rates for both $U_x$ and $U_y$ parameters. There is a remaining constant offset in $U_x$, but the differences between $U_y$ for the two rates are much smaller than those in the top row. Notably, the asymmetry is smaller too, and the regions with the larger $U_x$ offsets appear to have large $U_y$ differences as well as larger asymmetry around zero. 

\begin{figure}
\centering
\includegraphics[width=0.95\textwidth]{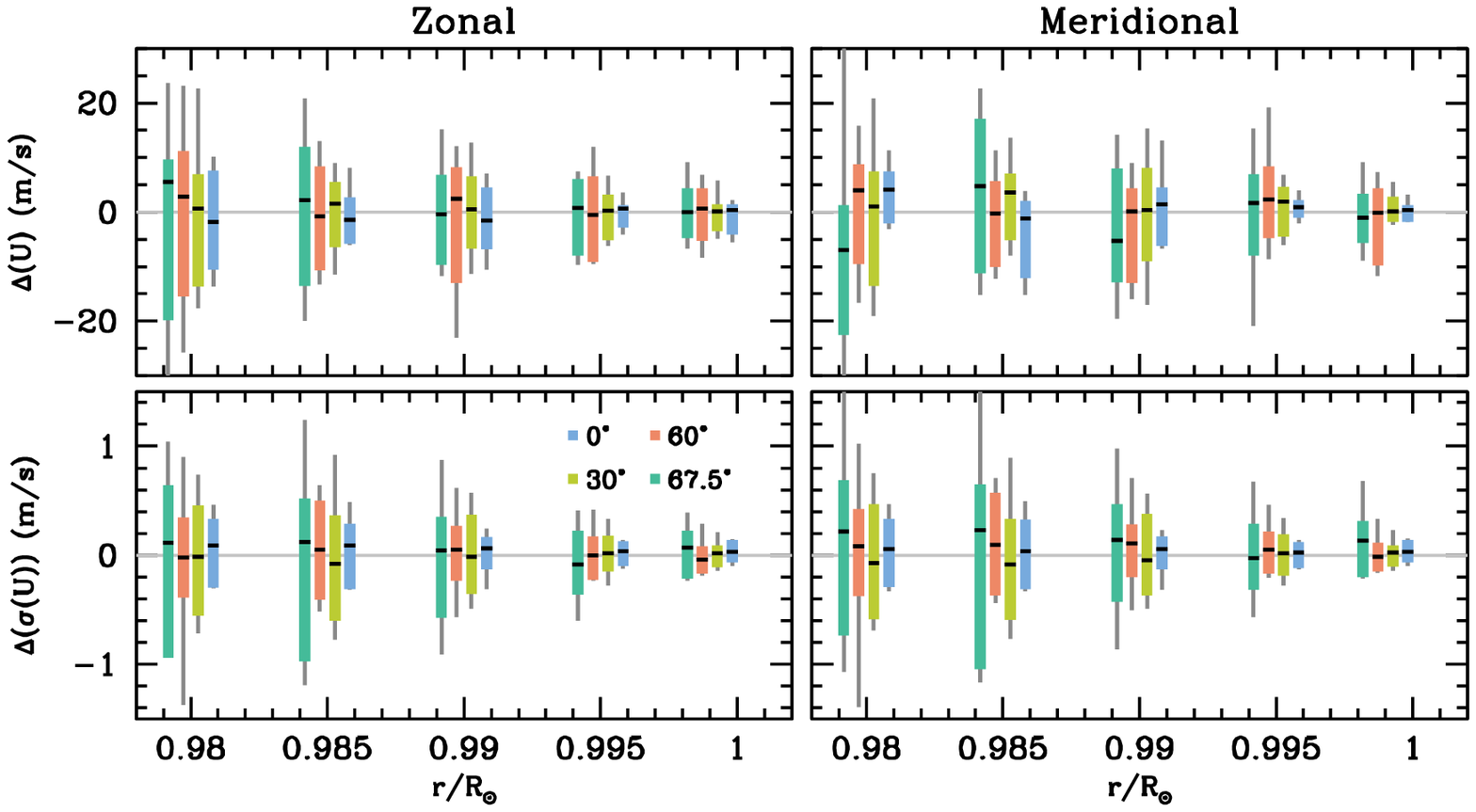}
\caption{Box-and-whisker plots showing the differences in inversion results (upper panels) and uncertainties (lower panels) obtained with Carrington and Snodgrass tracking. The zonal differences are actually $\Delta U_x-\langle\Delta U_x\rangle$, where $\langle\Delta U_x\rangle$ is the average at a given latitude and depth; this accounts for the large difference in the tracking rates. The results for each latitude (shown by different colors) are shifted slightly in radius to avoid overlapping results. Results are for inversions centered at 0.98, 0.985, 0.99, 0.995 and 0.999 R$_\odot$.
} 
\label{fig:track_carr}
\end{figure}
\begin{figure}
\centering
\includegraphics[width=0.95\textwidth]{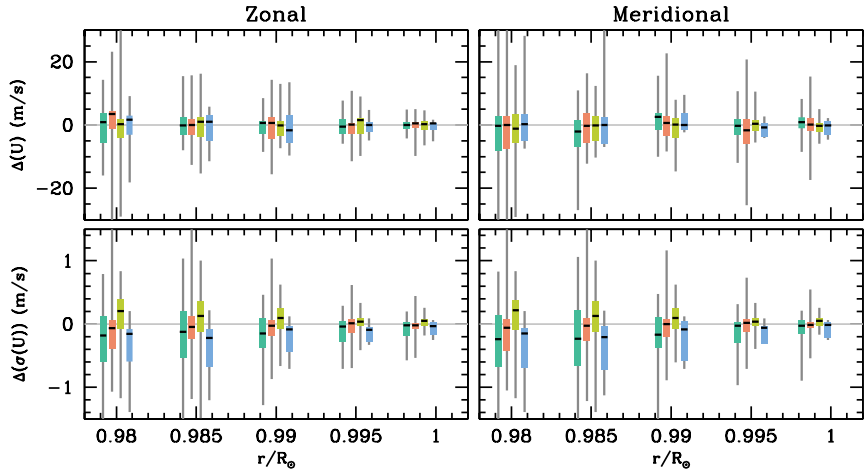}
\caption{The same as Fig.~\ref{fig:track_carr} but showing the differences between Snodgrass tracked and Helioseismic tracked results. In this case, the $U_x$ results have not been modified by subtracting the average difference.
} 
\label{fig:track_diff}
\end{figure}
\begin{figure}
\centering
	\includegraphics[width=0.95\textwidth]{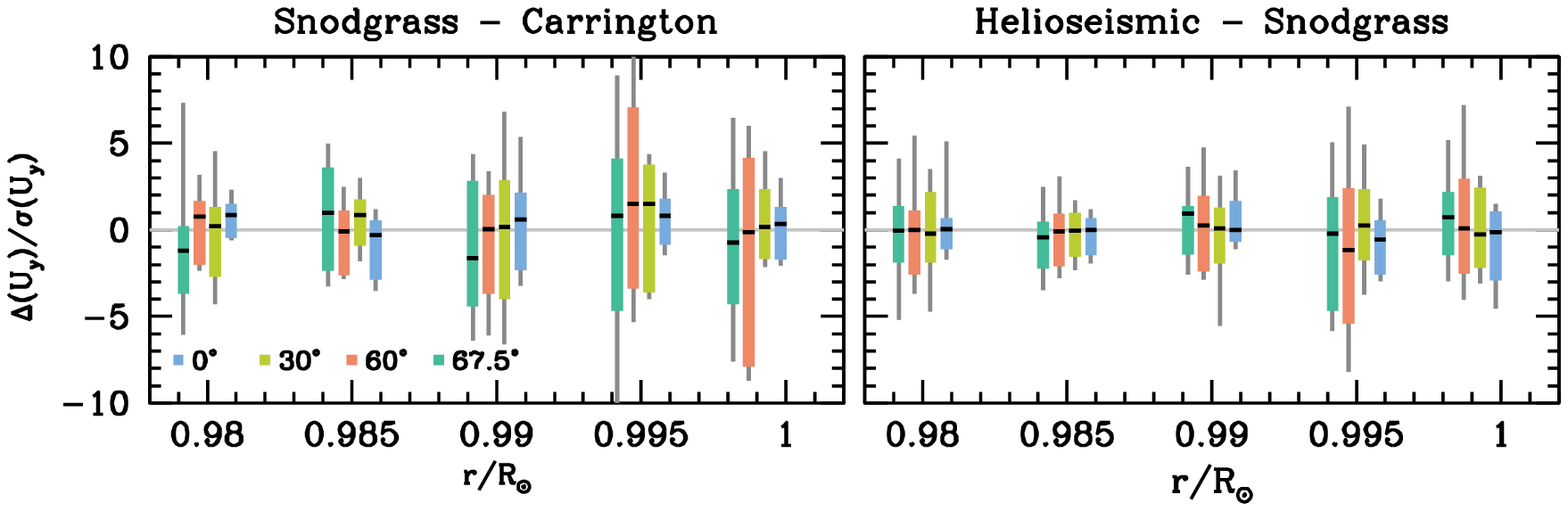}
\caption{{ Box-and-whisker diagrams of meridional flow differences caused by differences in the tracking rate normalized by uncertainties (i.e., the $z$-scores). The panel on the left shows Snodgrass$-$Carrington, the panel on the right shows results of Helioseismic$-$Snodgrass.}
} 
\label{fig:track_z}
\end{figure}

These results were obtained using fitted modes of all degrees and orders, and do not say anything about the depth dependence of the results. We inverted the results for each region on the central meridian and examined the differences as a function of depth and distance from the disc center. The results are shown in Figs.~\ref{fig:track_carr}\ and \ref{fig:track_diff}. The figures show so-called ``box and whiskers'' plots of the differences. 
The colored bars show the differences between the 16th and 84th quantile (if the distribution of differences were a Gaussian, these would correspond to $\pm 1 \sigma$), the horizontal line is the median, and the vertical lines show the region between the 2nd and the 98th quantile (i.e., approximately $\pm 2\sigma$). As one can see from Fig.~\ref{fig:track_diff}, the 1$\sigma$ spread in the results is much smaller (approximately $\pm 5$ m$s^{-1}$) when the tracking rates are similar; the $2\sigma$ spread in the results, however, is similar. Interestingly, but expected given the results in Fig.~\ref{fig:cdf_carr}, differences in the zonal tracking rate result in differences in the meridional flows. The differences are not inconsistent with the scatter in the results in Fig.~\ref{fig:fid}; moreover, the median difference in the results is close to zero, assuring us that the HMI flow results are reliable. 
{ To gauge whether or not the differences due to tracking are significant, in Fig.~\ref{fig:track_z} we show the distributions of the differences normalized by uncertainties. It is clear from the figure that the median change is generally less than 1$\sigma$, the effect of the tracking differences is seen in the size of the ``boxes'' which are supposed to be a $2\sigma$ spread, but for the (Snodgrass $-$ Carrington) case are sometimes greater than 5$\sigma$. Although we have not shown any {\tt rdfitf} results, we should note that the absolute differences are larger; because the uncertainties are larger too, however, they are not statistically significant.}

This does raise the question of what is the ``best'' rate to track regions --- different tracking rates essentially mean that we are not looking at exactly the same region. However,
given that the regions are tracked for relatively short intervals, the mismatch is small relative to the size of the region. And of course, the tracking rate can only be ``correct'' for a particular latitude and a particular depth. The extent of this effect may be understood by considering the situation at $60^\circ$ latitude, where our mean measured zonal flow (really the mean rotation at target depth in the Carrington frame) is its greatest, $\sim200$ m/s. Since regions are tracked for the length of time for them to rotate through their diameter, the fraction of the region which is ``misregistered'' due to differential rotation over the course of the tracking is the non-dimensionalized value of the differential rotation rate in units of $^\circ/^\circ$ in the Carrington frame, or about 4\% at that latitude. Likewise, the total linear distance traversed by a feature moving at the differential rotation rate at that depth and latitude over the course of a $15^\circ$ interval is about 20 Mm, roughly the size of a single supergranule.
Also, we cannot rule out a pure numerical effect in the fitting procedure; both \rdfitc\ and \rdfitf\ were developed for fitting MDI's Snodgrass-rate tracked rings, and the effect of the large distortion of the rings in the zonal direction caused by Carrington-rate tracking could potentially lead to cross-talk between the fitted parameters $U_x$ and $U_y$. In any case, the effect is only marginally significant { (see Fig.~\ref{fig:track_z})}. Nevertheless, this merits  additional study using a larger data sample, different tracking durations, as well as a thorough check of correlations among parameters obtained by the mode-fitting routines; such a study is beyond the scope of this work.

\subsection{Effects of resolution}
\label{subsec:resolution}

\begin{figure}
\centering
\includegraphics[width=1.0\textwidth]{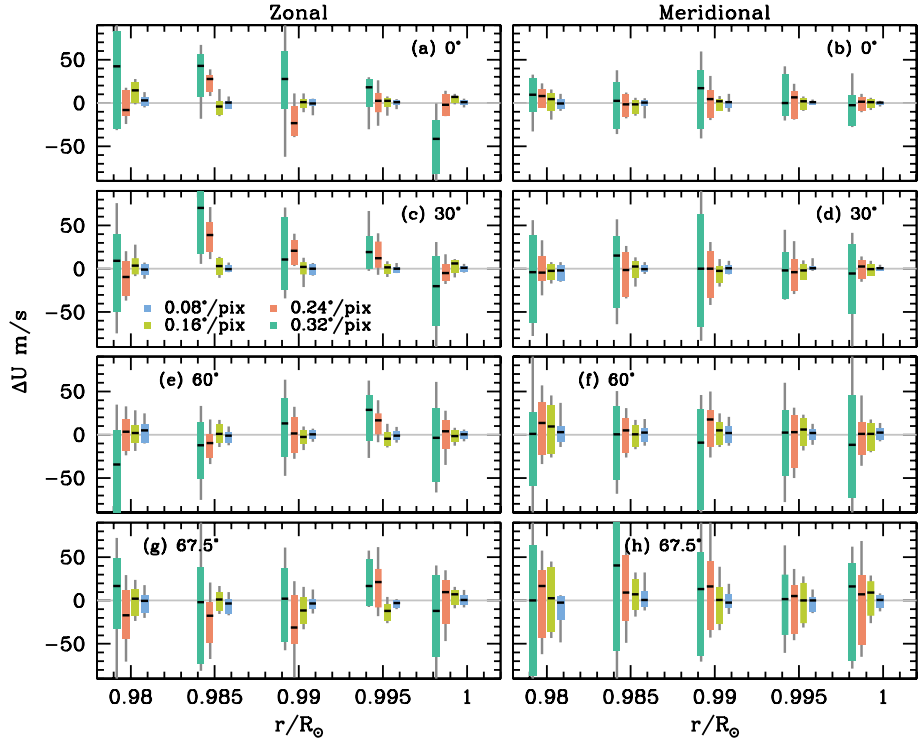}
\caption{Box-and-whisker diagram showing differences in inversion results when the spatial resolution of the Dopplergrams is changed. The differences are with respect to the fiducial HMI resolution of $0.04^\circ$ per pixel. Panels on the left show differences in zonal flow velocities, panels on the right show meridional flow differences. Colors indicate resolution.
} 
\label{fig:res_vel}
\end{figure}

\begin{figure}
\centering
\includegraphics[width=1.0\textwidth]{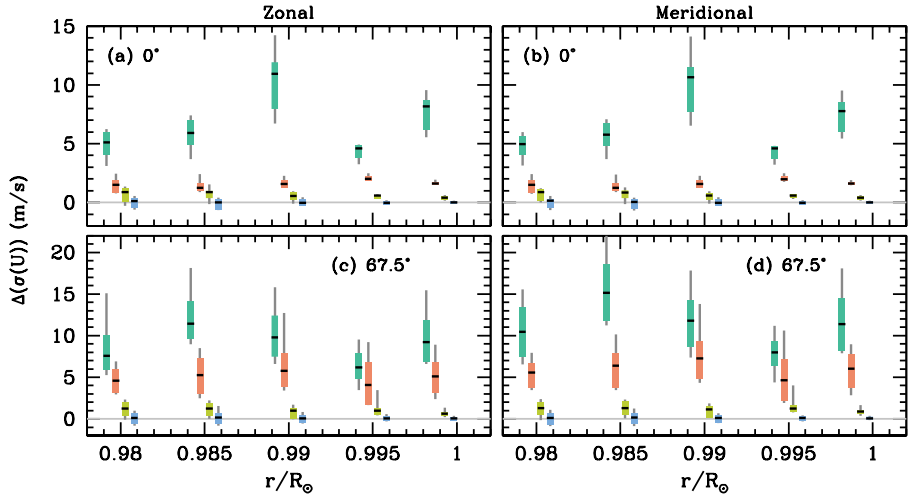}
\caption{The same as Fig.~\ref{fig:res_vel} but showing the differences in the uncertainties in the inversion results. We only show results for two latitudes on the meridian, $0^\circ$ (``disc center'') and the average of $\pm67.5^\circ$; the results for other latitudes lie in between these two extremes.}
\label{fig:res_err}
\end{figure}

We examine the effects of reduced spatial resolution next. The difference in $U_x$ and $U_y$ inversion results between rings created with different mapped pixel resolutions and the standard HMI one ($0.04^\circ$ heliographic  per pixel) are shown in Fig.~\ref{fig:res_vel}. As can be seen, the differences for results with $0.08^\circ$/pixel and $0.16^\circ$/pixel are quite small, especially at low latitudes, and the median differences for these cases is again close to zero. Recall that MDI rings are constructed with Dopplergrams with a resolution of $0.125^\circ$/pixel, thus resolution that is slightly worse than MDI does not appear to affect results significantly at low latitudes. The differences are larger at higher latitudes, and that is a direct effect of foreshortening being worse when the pixels are larger. This is why standard MDI ring diagrams were not constructed for latitudes higher than  $\pm 60^\circ$ while HMI ones are. The effects of resolution are most obvious when resolution is degraded further, and we can see large changes in the results for
$0.24^\circ$ and $0.32^\circ$/pixel. The median difference is often non-zero for these cases and the scatter is large. However, the  spread in the velocity differences is consistent with the uncertainties in the inversion results, which change with the pixel scale.  The differences in uncertainties in the inversion results are shown in Fig.~\ref{fig:res_err}. As can be seen, the uncertainties increase systematically as the resolution becomes worse.


 The change in uncertainties can be understood in terms of the modes, and the number of modes, that can be fitted. A large pixel scale means a larger ${\Delta}k$ --- the minimum difference in wavenumber $k$ --- which means that fewer modes can be fitted, and in turn less information. The larger pixels also mean that the Nyquist limit on the wavenumber is lower, so that one cannot obtain parameters for modes of very high degree, which adversely affects inversion results, particularly close to the surface.  { The change in uncertainties is particularly noticeable, particularly at 0.99\rsun. This is a result of the fact that these sets have (a) very few $f$-modes, (b) the $f$-modes have large errors, particularly the ones with the higher inertia, and (c) in this case the number of $n=1$ modes is small too. It just happens that the data limitations affect inversion results in the range $0.985$\rsun\ to about $0.993$\rsun.}
 
The parameters for the inversions were chosen to keep the radial resolution of the inversions similar in all cases, to allow us to compare the results at a given radius. Given that in inversion results radial resolution and statistical uncertainty have an inverse relationship --- the poorer the resolution, i.e, the larger the width of the averaging kernels, the lower the uncertainty ---  making the resolution kernels wider when inverting the lower resolution data will decrease the uncertainties. We could have opted to choose inversion parameters such that the uncertainties were similar, but in that case we would not be comparing the flow results around the same depth. 

{ It should be noted that the effect of degraded resolution on {\tt rdfitf}
results is much larger. We could not fit regions with 0.32$^\circ$/pixel resolution at all with \rdfitf. We only obtained a few tens of viable modes for regions with of $0.24^\circ$/pixel, not enough to be able to invert. For the other two cases, the absolute differences are larger, but consistent with the larger uncertainties}

\subsection{Effects of Noise}
\label{subsec:noise}

\begin{figure}
\centering
\includegraphics[width=0.85\textwidth]{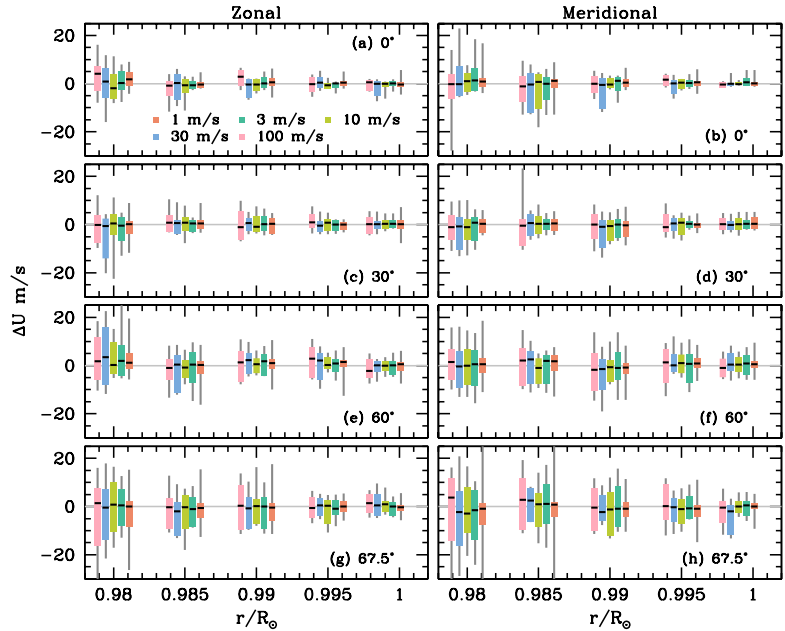}
\caption{Box-and-whisker diagrams for differences in inversion results between sets with added Doppler noise than the standard HMI set. Note that the y-scale is smaller than that in Fig.~\ref{fig:res_vel}, showing that greater Doppler noise has a minimal effect on the results, except at high latitudes.
} 
\label{fig:nois_vel}
\end{figure}

\begin{figure}
\centering
\includegraphics[width=0.85\textwidth]{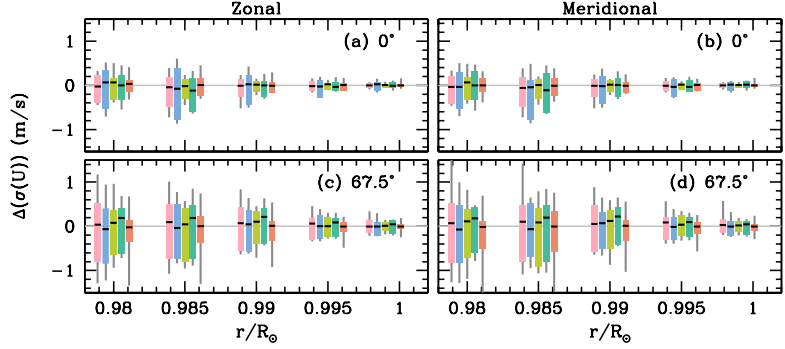}
\caption{The same as Fig.~\ref{fig:nois_vel} but showing the differences in the uncertainties in the inversion results. We show only the results for the same regions as in Fig.~\ref{fig:res_err}; the results for other latitudes lie in between these two extremes.
} 
\label{fig:nois_err}
\end{figure}

Added noise has a much smaller effect on the results; this is  
shown in Fig.~\ref{fig:nois_vel}. Note that the abscissa has a much smaller scale than that of Fig.~\ref{fig:res_vel} which shows the effect of reduced resolution. Changes in the uncertainties are shown in Fig.~\ref{fig:nois_err}.
 The differences in flow velocities at low latitudes are negligible close to the surface, even at the highest noise levels. At these latitudes, noise only affects the deepest layers; this is a result of the increasing difficulty of fitting the higher-order ($n$) modes that allow deeper inversions. However, the effect is still much smaller than that of degraded resolution. The effect on the uncertainties of the results are small too. Even at high latitudes, the median difference is close to zero at all depths, and the scatter in the results is consistent with the formal uncertainties in the inversion results.
 
 
{ As in the case where we examined the effect of using different tracking rates, results obtained with {\tt rdfitf} parameters show larger absolute changes; because of the large uncertainties in the results, however, the changes are not statistically significant.} 
 
\section{Effect of noise on 30-degree tiles} 
\label{sec:30deg}

\begin{figure}
\centering
\includegraphics[width=0.8\textwidth]{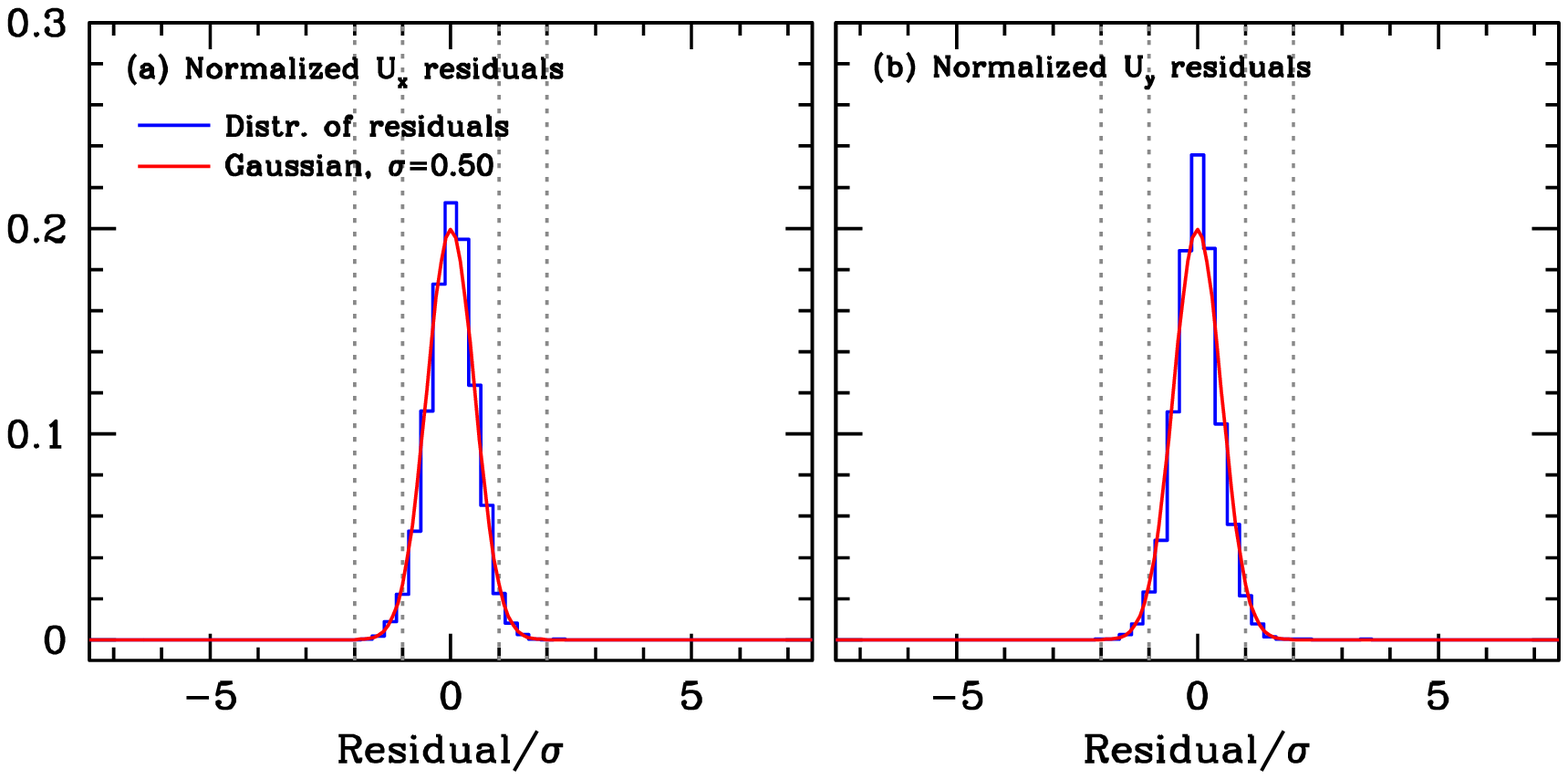}
\centering
\includegraphics[width=0.8\textwidth]{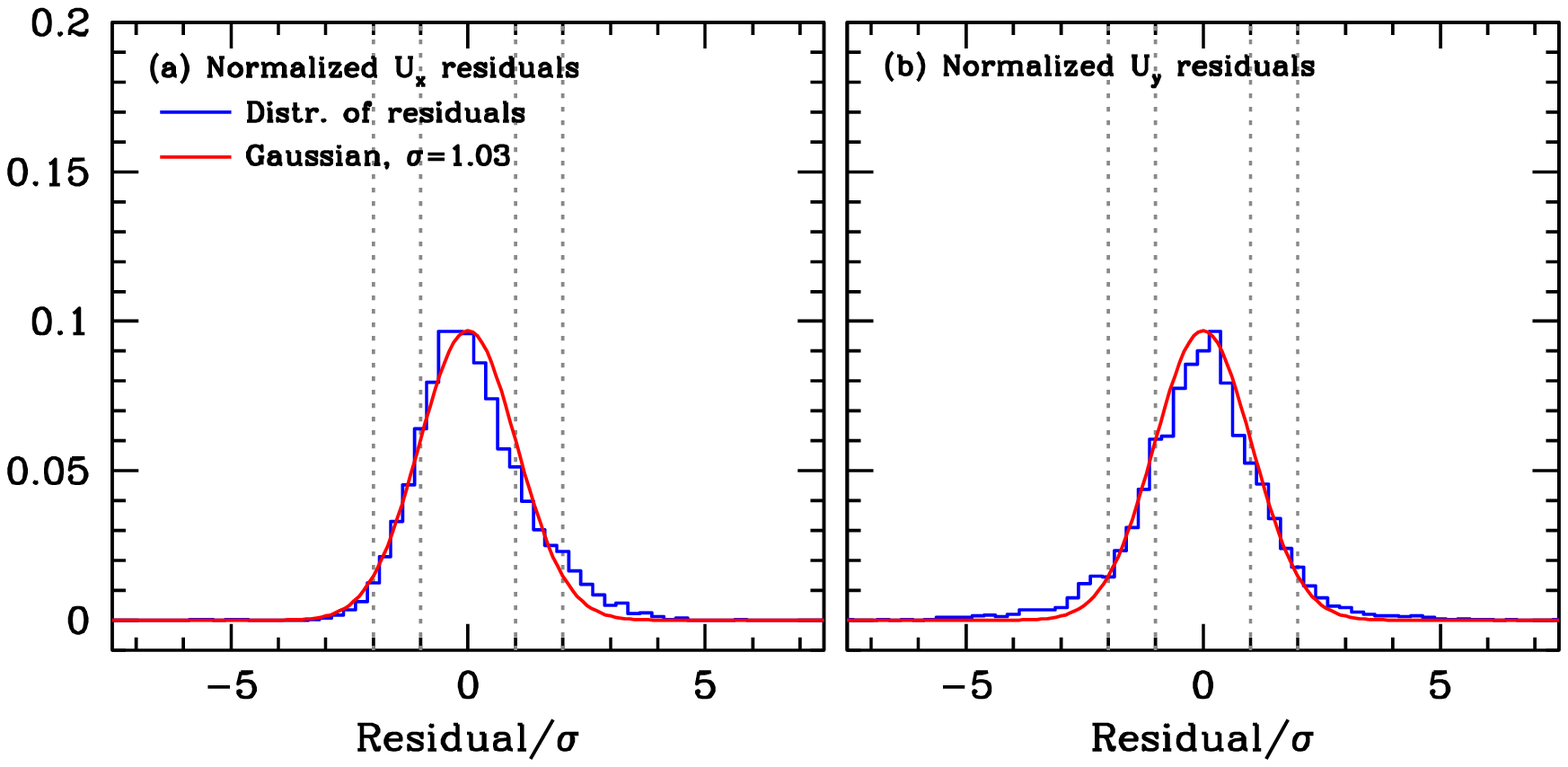}
\caption{{\bf Top row:}The distribution of residuals normalized by uncertainty for RLS inversions of {\rdfitf} parameters of all $30^\circ$ regions are shown as the blue histograms. The vertical dotted lines mark $\pm 1\sigma$ and $\pm 2\sigma$. The red curve shows a Gaussian
with $\sigma=0.50$ fitted to the histogram. {\bf Bottom row:} Residuals for rdfitc\ parameters. The red curve shows a Gaussian
with $\sigma=1.03$ fitted to the histogram.
In all panels, the vertical dotted lines mark $\pm 1\sigma$ and $\pm 2\sigma$.
} 
\label{fig:fitf_30res}
\end{figure}

Ring diagrams constructed with $30^\circ$ tiles allow us to probe the dynamics of the near-surface shear layer in its entirety. Inversion results are reliable from the surface down to about 0.95 R$_\odot$, and even deeper at the disc center. As in the case of the standard $15^\circ$ tiles, \rdfitf\ fits overestimate mode errors, and in this case by almost a factor of two; this can be seen from the distribution of residuals in the top row of Fig.~\ref{fig:fitf_30res}. Residuals from \rdfitc\ parameters seem to suggest that the uncertainties are reasonable (Fig.~\ref{fig:fitf_30res} bottom row).

\begin{figure}
\centering
\includegraphics[width=0.8\textwidth]{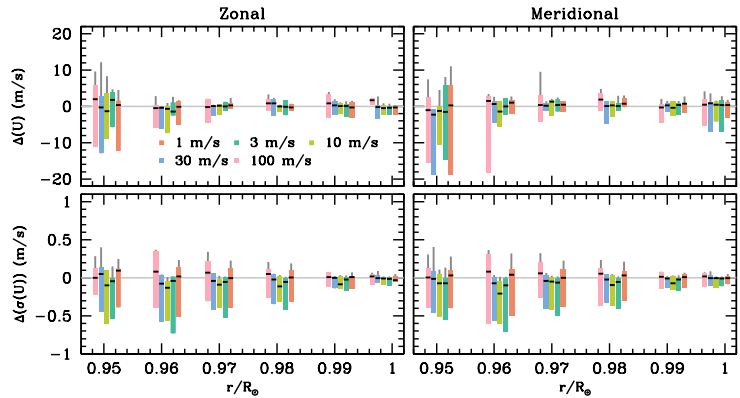}
\caption{ Box-and-whisker plots for differences between noise-added data and the standard dataset for $30^\circ$ tiles. Results are shown at 0.951, 0.961, 0.970, 0.981, 0.991 and 0.9985 R$_\odot$. 
The upper panels show differences in velocities, the lower panel shows differences in the estimated errors. The left panels are results for zonal velocities, the right ones are for meridional velocities. 
} 
\label{fig:fitc_30vel}
\end{figure}

The differences in the flow inversion results between the noise-added and the standard $30^\circ$ rings are shown in Fig.~\ref{fig:fitc_30vel}; the figure also shows the differences in the formal uncertainties in the flow-inversions. We show the combined results for the equator and latitude $\pm45^\circ$, since there are not many regions in our sample. The median differences in the inversion results are again close to zero, and the spread is consistent with the formal uncertainties. The changes are much smaller than for the $15^\circ$ case (Fig.~\ref{fig:nois_vel}); this is consistent with the fact that uncertainties in the $30^\circ$ parameters are less than those in the $15^\circ$ parameters. The largest effect is seen at the deepest layers that are inaccessible to $15^\circ$ rings. There is a larger change (compared with the $15^\circ$ case) at the shallowest layer too. That is undoubtedly because HMI $30^\circ$ rings have a lower spatial resolution (see Section~\ref{sec:data}). The changes in the uncertainties are also smaller than in the $15^\circ$ case. We were somewhat perplexed as to why the error-added data seems to have a large spread of cases with lower uncertainties, while the spread is smaller on the side with the (expected) larger errors; however, looking at the distribution of errors, we found that the effect is mostly caused by outliers, which have a greater influence because of the relatively small number of regions. These results lead us to conclude that added Doppler noise, at least to the level of 100 m/s, does not affect the inferred flow velocities. This is not very surprising, given the several times greater magnitude of the ``solar noise'' due to granulation and supergranulation.

\section{Results for 5-degree tiles} 
\label{sec:5deg}

\begin{figure}
\centering
\includegraphics[width=0.8\textwidth]{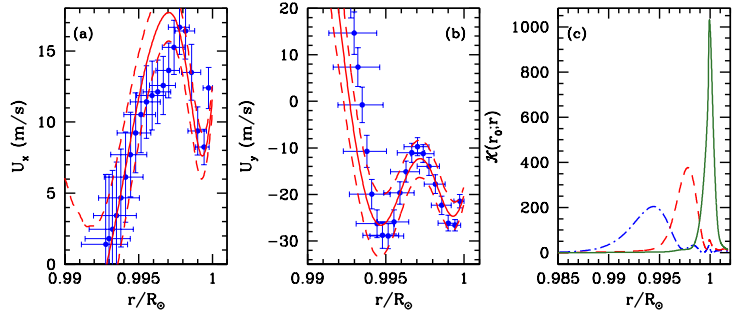}
\caption{(a) Zonal flow ($U_x$) inversion result for the $5^\circ$ HMI tile at disc center for CR 2224:005; { (b) meridional flow ($U_y$) inversion for the same region.} The blue points with error bars are OLA inversion  results, the red, continuous line is the RLS inversion result with 1$\sigma$ errors shown by the dashed lines. 
{ (c)} A few averaging kernels obtained for the OLA {  $U_x$} inversions. { Very similar averaging kernels are obtained for $U_y$ inversions, and hence are not shown separately.}
} 
\label{fig:inv_5ex}
\end{figure}

\begin{figure}
\centering
\includegraphics[width=0.8\textwidth]{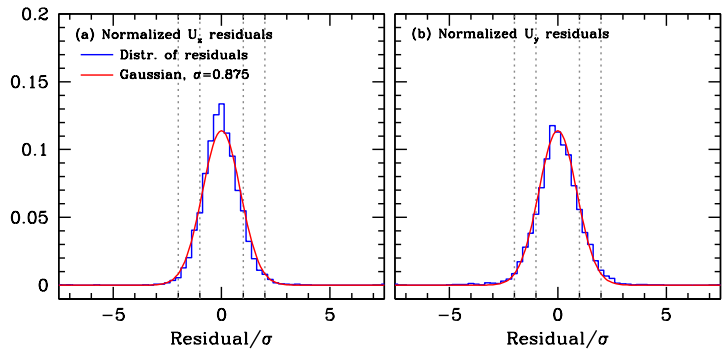}
\caption{The distribution of residuals normalized by uncertainty for RLS inversions of {\rdfitc} parameters of all $5^\circ$ regions are shown as the blue histograms. The vertical dotted lines mark $\pm 1\sigma$ and $\pm 2\sigma$. The red curve shows a Gaussian
with $\sigma=0.875$ fitted to the histogram, denoting that uncertainties in the flow parameters are slightly overestimated. 
} 
\label{fig:fitc_5res}
\end{figure}

\begin{figure}
\centering
\includegraphics[width=0.8\textwidth]{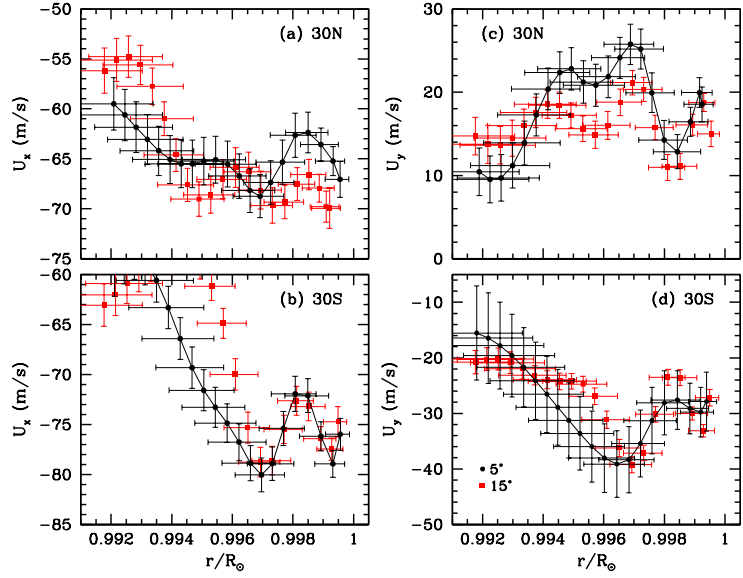}
\caption{ { A comparison of $15^\circ$ and $5^\circ$ inversion results for  $15^\circ$-tiles at latitudes of $\pm 30^\circ$ along the central meridian during CR 2224, CM longitude $60^\circ$. The points in red are the $15^\circ$ results, while the points in black are $5^\circ$ results obtained by averaging all $5^\circ$ regions in the given $15^\circ$ region (see text for details). }
} 
\label{fig:5degcomp}
\end{figure}

\begin{figure}
\centering
\includegraphics[width=0.9\textwidth]{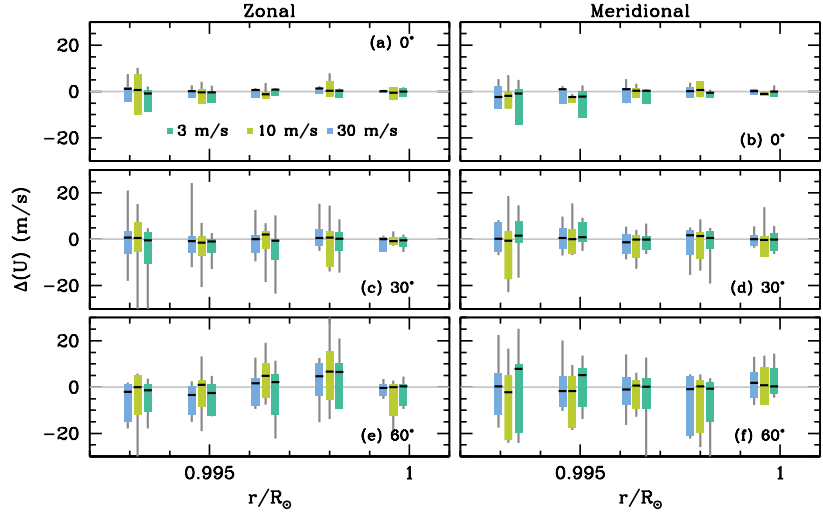}
\caption{Box-and-whisker plots for differences between noise-added data and the standard dataset for $5^\circ$ tiles. Results are shown at 0.9932, 0.9948, 0.9964, 0.9980 and 0.9996 $R_\odot$. As usual, the results for the different
noise cases at any given radius have been displaced slightly for visibility. Each group, from right to left, shows the difference in results for added noise of 3 ms$^{-1}$, 10 ms$^{-1}$ and 30 ms$^{-1}$.
} 
\label{fig:fitc_5box}
\end{figure}
\begin{figure}
\centering
\includegraphics[width=0.9\textwidth]{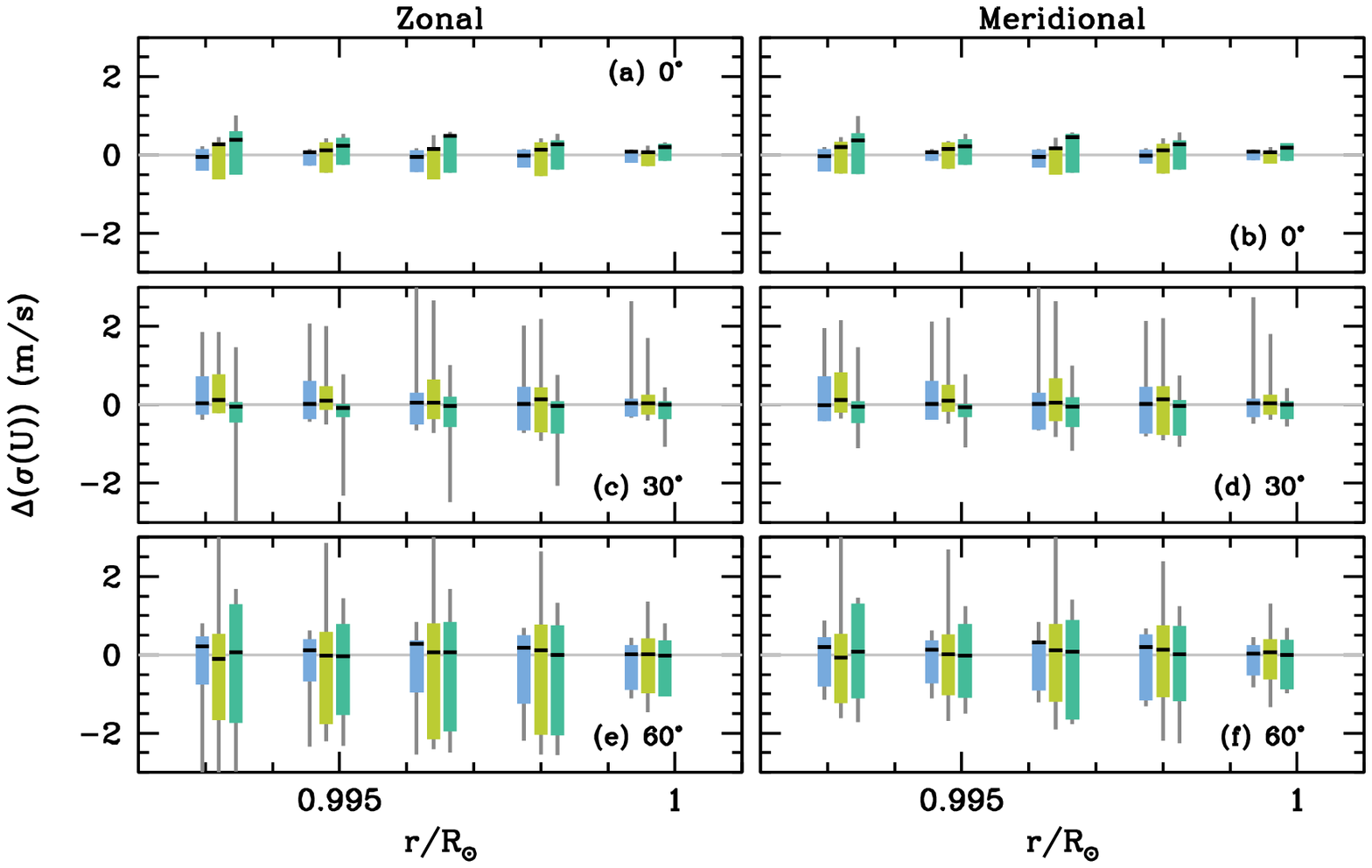}
\caption{The same as Fig.~\ref{fig:fitc_5box}, but showing the change in the uncertainties in the inversion results.
} 
\label{fig:fitc_5boxsig}
\end{figure}

While the HMI pipeline produces $5^\circ$ tiles routinely, and also provides fitted parameters (only with \rdfitc), it does not produce inversion results; it was always assumed that inversions of the fitted modes would be difficult, if not impossible, because only parameters for a small handful of modes can be determined, { a few hundred modes, instead of one thousand or more for $15^\circ$ rings}. Most of the fitted modes are the $f$, $p_1$ and $p_2$ modes, { though it is possible to fit the $n=3$ modes for many regions.}. These modes have been used to study near-surface flows \citep[e.g.][]{persistent} assuming that averaging the flow parameters in bins of different lower-turning points of the modes allows us to estimate the flow velocities. It should be noted that even smaller rings have been used in inversions \citep{3drings}, but only in combination with much larger rings, and not on their own.
As we show in Fig.~\ref{fig:inv_5ex}, the $5^\circ$ rings can be inverted in isolation. The OLA and RLS results agree reasonably well, and the OLA averaging kernels are well localized. As is seen from the figure, we obtain independent velocity estimates at two { depths}. { Inversions are possible at high latitudes also; although in this study we have not used $5^\circ$ rings at very high latitudes, a judicious selection of inversion parameters allows us to invert  regions above $70^\circ$; the limiting latitude is being investigated.} RLS residuals imply that the mode uncertainties are overestimated by \rdfitc\ by about 15\% (see Fig.~\ref{fig:fitc_5res}).

{
It could be asked how the $5^\circ$ results compare with $15^\circ$ results. Such a comparison, however, is not straightforward. For one, flows change over short distances, the $15^\circ$ tiles cover many $5^\circ$ tiles. If ring diagrams used square tiles, this would not be a problem, there would be nine $5^\circ$ tiles per $15^\circ$ tile, but the regions in both cases are circular. Another issue is that 15-degree tiles are tracked three times longer than 5-deg tiles, and thus have three times better frequency resolution (and three times poorer temporal resolution!). To compare $5^\circ$ and $15^\circ$ results we worked around these problems by (a) covering the 15-degree tiles by the 27 $5^\circ$ tracked tiles over the longer interval whose circular apodizations were more than half within that of the $15^\circ$ tile, to ensure coverage in both space and time; (b) averaging the mode parameters obtained using the 27 $5^\circ$ tiles; (c) inverting the 15-degree tiles using only every 3rd mode to account for the difference in frequency resolution; and (d) not using the $n \ge 3$ modes, since those cannot be reliably fitted for 5-degree tiles. We show the results in Fig.~\ref{fig:5degcomp}.
As the results show, the flow structure determined from the average parameters obtained from the 27 $5^\circ$ rings that cover a standard $15^\circ$ ring agrees well with that found with the $15^\circ$ ring data.
}

Added Doppler noise has a slightly larger effect on $5^\circ$ flow results than for the larger rings; however, the median change is again consistent with zero (Fig.~\ref{fig:fitc_5box}). The spread in values is higher at higher latitudes, but they are consistent with the uncertainties in the results. The change in uncertainties is larger than  for the $15^\circ$ and the $30^\circ$ cases (see Fig.~\ref{fig:fitc_5boxsig}), but the changes are not large enough to make the inversion results unreliable.

\section{Summary}
\label{sec:disc}

We have conducted a study to examine the effects of tracking rate, spatial resolution and Doppler noise on results obtained by ring-diagram analysis of helioseismic data. We use the HMI ring-diagram pipeline to carry out the analysis. 

Our results show that for both $15^\circ$ and $30^\circ$ rings, uncertainties returned by \rdfitf\ are overestimated. This would imply that inversion results provided by the HMI project are not as precise as they could be, given that those are inversions of parameters obtained by \rdfitf. { Given this, in addition to our finding that the $5^\circ$ \rdfitc\ fits produced by the HMI pipeline can be inverted, it appears desirable to add
inversions of all of the \rdfitc\ fits to the set of HMI pipeline products. That this was not done in the first place is that due to the computational cost ---  during the first five years of the SDO mission, \rdfitc\ fits to $15^\circ$ and $30^\circ$ power spectra were only performed for tiles along the central meridian and equator. With additional computing resources, full-disc fits were added to the pipeline beginning with Carrington Rotation 2172 (Dec. 2015); however, retrospective processing of the data for the full mission has only very recently been completed.}

We find that the rate at which regions on the solar surface are tracked in the zonal direction has a small effect on the flow parameters obtained in the meridional direction. However, the differences are small, well within statistical uncertainties, and the overall effect on the inversion results is small as well.

The largest effect on flow results is that of spatial resolution: small pixels give better results than larger ones. If we keep the radial resolution of the inversions the same for all pixel scales, the uncertainties in the inversion results increase substantially. That said, pixel scales of up to $0.16^\circ$ per pixel (i.e., somewhat poorer resolution than MDI) give similar results. 

Added Doppler noise has almost no effect on the flows. While this might seem surprising, it is not if one considers the magnitude of the Doppler signal in the modes themselves, as well as the magnitude of the noise due to granulation and supergranulation. The effect is even smaller if one examines the results obtained with $30^\circ$ tiles.

We find that it is possible to obtain reliable inversion results from $5^\circ$ rings alone. This opens up the possibility of studying high-latitude flows --- $5^\circ$
rings with HMI data can be constructed at latitudes as high as $80^\circ$ depending on the value of the $B_0$ angle. The flow parameters obtained are somewhat more sensitive to noise than the larger tiles; even then, however, the sensitivity is not so great as to make flow estimates unreliable.

\section*{Data Availability Statement:}
The HMI data used in this article are publicly available from the JSOC at jsoc.stanford.edu. The fitted parameters of the rings diagrams and inversion results used for the current study are available from the corresponding author on reasonable request.

\begin{acks}
This work uses data from the Helioseismic and Magnetic Imager. HMI data are courtesy of NASA/SDO and the HMI science team. This research was supported in part by NASA Contract NAS5-02139 to Stanford University.
\end{acks}

     

\begin{thebibliography}{25}
\ifx\bisbn     \undefined \def\bisbn  #1{ISBN #1}\fi
\ifx\binits    \undefined \def\binits#1{#1}\fi
\ifx\bauthor   \undefined \def\bauthor#1{#1}\fi
\ifx\batitle   \undefined \def\batitle#1{#1}\fi
\ifx\bjtitle   \undefined \def\bjtitle#1{\textit{#1}}\fi
\ifx\bvolume   \undefined \def\bvolume#1{\textbf{#1}}\fi
\ifx\byear     \undefined \def\byear#1{#1}\fi
\ifx\bissue    \undefined \def\bissue#1{#1}\fi
\ifx\bfpage    \undefined \def\bfpage#1{#1}\fi
\ifx\blpage    \undefined \def\blpage #1{#1}\fi
\ifx\burl      \undefined \def\burl#1{\textsf{#1}}\fi
\ifx\href      \undefined \def\href#1#2{\textsf{#2}}\fi
\ifx\betal     \undefined \def\betal{\textit{et al.}}\fi
\ifx\bctitle   \undefined \def\bctitle#1{#1}\fi
\ifx\beditor   \undefined \def\beditor#1{#1}\fi
\ifx\bbtitle   \undefined \def\bbtitle#1{\textit{#1}}\fi
\ifx\bedition  \undefined \def\bedition#1{#1}\fi
\ifx\bseriesno \undefined \def\bseriesno#1{\textbf{#1}}\fi
\ifx\blocation \undefined \def\blocation#1{#1}\fi
\ifx\bsertitle \undefined \def\bsertitle#1{\textit{#1}}\fi
\ifx\bsnm      \undefined \def\bsnm#1{#1}\fi
\ifx\bsuffix   \undefined \def\bsuffix#1{#1}\fi
\ifx\bparticle \undefined \def\bparticle#1{#1}\fi
\ifx\barticle  \undefined \def\barticle#1{}\fi
\ifx\binstitute  \undefined \def\binstitute#1{#1}\fi
\ifx\bpublisher  \undefined \def\bpublisher#1{#1}\fi
\ifx\doiurl    \undefined
  \def\doiurl#1{\href{http://dx.doi.org/#1}{\textsf{DOI}}}\fi
\ifx\arxivurl  \undefined
  \def\arxivurl#1{\href{http://arxiv.org/abs/#1}{\textsf{arXiv}}}\fi
\ifx\adsurl    \undefined
  \def\adsurl#1{\href{http://adsabs.harvard.edu/abs/#1}{\textsf{ADS}}}\fi
\ifx\botherref \undefined \def\botherref#1{}\fi
\ifx\url       \undefined \def\url#1{\textsf{#1}}\fi
\ifx\bchapter  \undefined \def\bchapter#1{}\fi
\ifx\bbook     \undefined \def\bbook#1{}\fi
\ifx\bcomment  \undefined \def\bcomment#1{#1}\fi
\ifx\oauthor   \undefined \def\oauthor#1{#1}\fi
\ifx\citeauthoryear \undefined\def \citeauthoryear#1{#1}\fi
\ifx\endbibitem\undefined \def\endbibitem{}\fi
\ifx\bconflocation  \undefined \def\bconflocation#1{#1} \fi

\bibitem[\protect\citeauthoryear{{Baldner}, {Bogart}, and
  {Basu}}{2013}]{baldner2013}
\begin{barticle}
\bauthor{\bsnm{{Baldner}}, \binits{C.S.}},
\bauthor{\bsnm{{Bogart}}, \binits{R.S.}},
\bauthor{\bsnm{{Basu}}, \binits{S.}}:
\byear{2013},
\batitle{{The Sub-surface Structure of a Large Sample of Active Regions}}.
\bjtitle{\solphys}
\bvolume{287}(\bissue{1-2}),
\bfpage{265}.
\doiurl{10.1007/s11207-012-0148-9}.
\adsurl{https://ui.adsabs.harvard.edu/abs/2013SoPh..287..265B}.
\end{barticle}
\endbibitem

\bibitem[\protect\citeauthoryear{{Basu} and {Antia}}{1999}]{rdfitc}
\begin{barticle}
\bauthor{\bsnm{{Basu}}, \binits{S.}},
\bauthor{\bsnm{{Antia}}, \binits{H.M.}}:
\byear{1999},
\batitle{{Large-Scale Flows in the Solar Interior: Effect of Asymmetry in Peak
  Profiles}}.
\bjtitle{\apj}
\bvolume{525}(\bissue{1}),
\bfpage{517}.
\doiurl{10.1086/307900}.
\adsurl{https://ui.adsabs.harvard.edu/abs/1999ApJ...525..517B}.
\end{barticle}
\endbibitem

\bibitem[\protect\citeauthoryear{{Basu} and {Antia}}{2010}]{basu2010}
\begin{barticle}
\bauthor{\bsnm{{Basu}}, \binits{S.}},
\bauthor{\bsnm{{Antia}}, \binits{H.M.}}:
\byear{2010},
\batitle{{Characteristics of Solar Meridional Flows during Solar Cycle 23}}.
\bjtitle{\apj}
\bvolume{717}(\bissue{1}),
\bfpage{488}.
\doiurl{10.1088/0004-637X/717/1/488}.
\adsurl{https://ui.adsabs.harvard.edu/abs/2010ApJ...717..488B}.
\end{barticle}
\endbibitem

\bibitem[\protect\citeauthoryear{{Basu}, {Antia}, and
  {Bogart}}{2004}]{basu2004}
\begin{barticle}
\bauthor{\bsnm{{Basu}}, \binits{S.}},
\bauthor{\bsnm{{Antia}}, \binits{H.M.}},
\bauthor{\bsnm{{Bogart}}, \binits{R.S.}}:
\byear{2004},
\batitle{{Ring-Diagram Analysis of the Structure of Solar Active Regions}}.
\bjtitle{\apj}
\bvolume{610}(\bissue{2}),
\bfpage{1157}.
\doiurl{10.1086/421843}.
\adsurl{https://ui.adsabs.harvard.edu/abs/2004ApJ...610.1157B}.
\end{barticle}
\endbibitem

\bibitem[\protect\citeauthoryear{{Basu}, {Antia}, and
  {Bogart}}{2007}]{basu2007}
\begin{barticle}
\bauthor{\bsnm{{Basu}}, \binits{S.}},
\bauthor{\bsnm{{Antia}}, \binits{H.M.}},
\bauthor{\bsnm{{Bogart}}, \binits{R.S.}}:
\byear{2007},
\batitle{{Structure of the Near-Surface Layers of the Sun: Asphericity and Time
  Variation}}.
\bjtitle{\apj}
\bvolume{654}(\bissue{2}),
\bfpage{1146}.
\doiurl{10.1086/509251}.
\adsurl{https://ui.adsabs.harvard.edu/abs/2007ApJ...654.1146B}.
\end{barticle}
\endbibitem

\bibitem[\protect\citeauthoryear{{Basu}, {Antia}, and
  {Tripathy}}{1999}]{ringsinv}
\begin{barticle}
\bauthor{\bsnm{{Basu}}, \binits{S.}},
\bauthor{\bsnm{{Antia}}, \binits{H.M.}},
\bauthor{\bsnm{{Tripathy}}, \binits{S.C.}}:
\byear{1999},
\batitle{{Ring Diagram Analysis of Near-Surface Flows in the Sun}}.
\bjtitle{\apj}
\bvolume{512}(\bissue{1}),
\bfpage{458}.
\doiurl{10.1086/306765}.
\adsurl{https://ui.adsabs.harvard.edu/abs/1999ApJ...512..458B}.
\end{barticle}
\endbibitem

\bibitem[\protect\citeauthoryear{{Bogart}, {Baldner}, and
  {Basu}}{2015}]{persistent}
\begin{barticle}
\bauthor{\bsnm{{Bogart}}, \binits{R.S.}},
\bauthor{\bsnm{{Baldner}}, \binits{C.S.}},
\bauthor{\bsnm{{Basu}}, \binits{S.}}:
\byear{2015},
\batitle{{Evolution of Near-surface Flows Inferred from High-resolution
  Ring-diagram Analysis}}.
\bjtitle{\apj}
\bvolume{807}(\bissue{2}),
\bfpage{125}.
\doiurl{10.1088/0004-637X/807/2/125}.
\adsurl{https://ui.adsabs.harvard.edu/abs/2015ApJ...807..125B}.
\end{barticle}
\endbibitem

\bibitem[\protect\citeauthoryear{{Bogart} \textit{et~al.}}{2008}]{bogart2008}
\begin{barticle}
\bauthor{\bsnm{{Bogart}}, \binits{R.S.}},
\bauthor{\bsnm{{Basu}}, \binits{S.}},
\bauthor{\bsnm{{Rabello-Soares}}, \binits{M.C.}},
\bauthor{\bsnm{{Antia}}, \binits{H.M.}}:
\byear{2008},
\batitle{{Probing the Subsurface Structures of Active Regions with Ring-Diagram
  Analysis}}.
\bjtitle{\solphys}
\bvolume{251}(\bissue{1-2}),
\bfpage{439}.
\doiurl{10.1007/s11207-008-9213-9}.
\adsurl{https://ui.adsabs.harvard.edu/abs/2008SoPh..251..439B}.
\end{barticle}
\endbibitem

\bibitem[\protect\citeauthoryear{{Bogart} \textit{et~al.}}{2011a}]{ringpipe1}
\begin{bchapter}
\bauthor{\bsnm{{Bogart}}, \binits{R.S.}},
\bauthor{\bsnm{{Baldner}}, \binits{C.}},
\bauthor{\bsnm{{Basu}}, \binits{S.}},
\bauthor{\bsnm{{Haber}}, \binits{D.A.}},
\bauthor{\bsnm{{Rabello-Soares}}, \binits{M.C.}}:
\byear{2011}a,
\bctitle{{HMI ring diagram analysis I. The processing pipeline}}.
In: \bbtitle{GONG-SoHO 24: A New Era of Seismology of the Sun and Solar-Like
  Stars},
\bsertitle{Journal of Physics Conference Series}
\bseriesno{271},
\bfpage{012008}.
\doiurl{10.1088/1742-6596/271/1/012008}.
\adsurl{https://ui.adsabs.harvard.edu/abs/2011JPhCS.271a2008B}.
\end{bchapter}
\endbibitem

\bibitem[\protect\citeauthoryear{{Bogart} \textit{et~al.}}{2011b}]{ringpipe2}
\begin{bchapter}
\bauthor{\bsnm{{Bogart}}, \binits{R.S.}},
\bauthor{\bsnm{{Baldner}}, \binits{C.}},
\bauthor{\bsnm{{Basu}}, \binits{S.}},
\bauthor{\bsnm{{Haber}}, \binits{D.A.}},
\bauthor{\bsnm{{Rabello-Soares}}, \binits{M.C.}}:
\byear{2011}b,
\bctitle{{HMI ring diagram analysis II. Data products}}.
In: \bbtitle{GONG-SoHO 24: A New Era of Seismology of the Sun and Solar-Like
  Stars},
\bsertitle{Journal of Physics Conference Series}
\bseriesno{271},
\bfpage{012009}.
\doiurl{10.1088/1742-6596/271/1/012009}.
\adsurl{https://ui.adsabs.harvard.edu/abs/2011JPhCS.271a2009B}.
\end{bchapter}
\endbibitem

\bibitem[\protect\citeauthoryear{{Featherstone}, {Hindman}, and
  {Thompson}}{2011}]{3drings}
\begin{bchapter}
\bauthor{\bsnm{{Featherstone}}, \binits{N.A.}},
\bauthor{\bsnm{{Hindman}}, \binits{B.W.}},
\bauthor{\bsnm{{Thompson}}, \binits{M.J.}}:
\byear{2011},
\bctitle{{Ring-analysis flow measurements of sunspot outflows}}.
In: \bbtitle{GONG-SoHO 24: A New Era of Seismology of the Sun and Solar-Like
  Stars},
\bsertitle{Journal of Physics Conference Series}
\bseriesno{271},
\bfpage{012002}.
\doiurl{10.1088/1742-6596/271/1/012002}.
\adsurl{https://ui.adsabs.harvard.edu/abs/2011JPhCS.271a2002F}.
\end{bchapter}
\endbibitem

\bibitem[\protect\citeauthoryear{{Gonz{\'a}lez Hern{\'a}ndez}
  \textit{et~al.}}{2006}]{irene2006}
\begin{barticle}
\bauthor{\bsnm{{Gonz{\'a}lez Hern{\'a}ndez}}, \binits{I.}},
\bauthor{\bsnm{{Komm}}, \binits{R.}},
\bauthor{\bsnm{{Hill}}, \binits{F.}},
\bauthor{\bsnm{{Howe}}, \binits{R.}},
\bauthor{\bsnm{{Corbard}}, \binits{T.}},
\bauthor{\bsnm{{Haber}}, \binits{D.A.}}:
\byear{2006},
\batitle{{Meridional Circulation Variability from Large-Aperture Ring-Diagram
  Analysis of Global Oscillation Network Group and Michelson Doppler Imager
  Data}}.
\bjtitle{\apj}
\bvolume{638}(\bissue{1}),
\bfpage{576}.
\doiurl{10.1086/498642}.
\adsurl{https://ui.adsabs.harvard.edu/abs/2006ApJ...638..576G}.
\end{barticle}
\endbibitem

\bibitem[\protect\citeauthoryear{{Haber} \textit{et~al.}}{2000a}]{haber2000}
\begin{barticle}
\bauthor{\bsnm{{Haber}}, \binits{D.A.}},
\bauthor{\bsnm{{Hindman}}, \binits{B.W.}},
\bauthor{\bsnm{{Toomre}}, \binits{J.}},
\bauthor{\bsnm{{Bogart}}, \binits{R.S.}},
\bauthor{\bsnm{{Thompson}}, \binits{M.J.}},
\bauthor{\bsnm{{Hill}}, \binits{F.}}:
\byear{2000}a,
\batitle{{Solar shear flows deduced from helioseismic dense-pack samplings of
  ring diagrams}}.
\bjtitle{\solphys}
\bvolume{192},
\bfpage{335}.
\doiurl{10.1023/A:1005235610132}.
\adsurl{https://ui.adsabs.harvard.edu/abs/2000SoPh..192..335H}.
\end{barticle}
\endbibitem

\bibitem[\protect\citeauthoryear{{Haber} \textit{et~al.}}{2000b}]{rdfitf}
\begin{barticle}
\bauthor{\bsnm{{Haber}}, \binits{D.A.}},
\bauthor{\bsnm{{Hindman}}, \binits{B.W.}},
\bauthor{\bsnm{{Toomre}}, \binits{J.}},
\bauthor{\bsnm{{Bogart}}, \binits{R.S.}},
\bauthor{\bsnm{{Thompson}}, \binits{M.J.}},
\bauthor{\bsnm{{Hill}}, \binits{F.}}:
\byear{2000}b,
\batitle{{Solar shear flows deduced from helioseismic dense-pack samplings of
  ring diagrams}}.
\bjtitle{\solphys}
\bvolume{192},
\bfpage{335}.
\doiurl{10.1023/A:1005235610132}.
\adsurl{https://ui.adsabs.harvard.edu/abs/2000SoPh..192..335H}.
\end{barticle}
\endbibitem

\bibitem[\protect\citeauthoryear{{Hanson}, {Gizon}, and
  {Liang}}{2020}]{hanson2020}
\begin{barticle}
\bauthor{\bsnm{{Hanson}}, \binits{C.S.}},
\bauthor{\bsnm{{Gizon}}, \binits{L.}},
\bauthor{\bsnm{{Liang}}, \binits{Z.-C.}}:
\byear{2020},
\batitle{{Solar Rossby waves observed in GONG++ ring-diagram flow maps}}.
\bjtitle{\aap}
\bvolume{635},
\bfpage{A109}.
\doiurl{10.1051/0004-6361/201937321}.
\adsurl{https://ui.adsabs.harvard.edu/abs/2020A&A...635A.109H}.
\end{barticle}
\endbibitem

\bibitem[\protect\citeauthoryear{{Hill}}{1988}]{frankhill}
\begin{barticle}
\bauthor{\bsnm{{Hill}}, \binits{F.}}:
\byear{1988},
\batitle{{Rings and Trumpets---Three-dimensional Power Spectra of Solar
  Oscillations}}.
\bjtitle{\apj}
\bvolume{333},
\bfpage{996}.
\doiurl{10.1086/166807}.
\adsurl{https://ui.adsabs.harvard.edu/abs/1988ApJ...333..996H}.
\end{barticle}
\endbibitem

\bibitem[\protect\citeauthoryear{{Hill} \textit{et~al.}}{1996}]{gong}
\begin{barticle}
\bauthor{\bsnm{{Hill}}, \binits{F.}},
\bauthor{\bsnm{{Stark}}, \binits{P.B.}},
\bauthor{\bsnm{{Stebbins}}, \binits{R.T.}},
\bauthor{\bsnm{{Anderson}}, \binits{E.R.}},
\bauthor{\bsnm{{Antia}}, \binits{H.M.}},
\bauthor{\bsnm{{Brown}}, \binits{T.M.}},
\bauthor{\bsnm{{Duvall}}, \binits{T.L.} \bsuffix{Jr.}},
\bauthor{\bsnm{{Haber}}, \binits{D.A.}},
\bauthor{\bsnm{{Harvey}}, \binits{J.W.}},
\bauthor{\bsnm{{Hathaway}}, \binits{D.H.}},
\bauthor{\bsnm{{Howe}}, \binits{R.}},
\bauthor{\bsnm{{Hubbard}}, \binits{R.P.}},
\bauthor{\bsnm{{Jones}}, \binits{H.P.}},
\bauthor{\bsnm{{Kennedy}}, \binits{J.R.}},
\bauthor{\bsnm{{Korzennik}}, \binits{S.G.}},
\bauthor{\bsnm{{Kosovichev}}, \binits{A.G.}},
\bauthor{\bsnm{{Leibacher}}, \binits{J.W.}},
\bauthor{\bsnm{{Libbrecht}}, \binits{K.G.}},
\bauthor{\bsnm{{Pintar}}, \binits{J.A.}},
\bauthor{\bsnm{{Rhodes}}, \binits{E.J.} \bsuffix{Jr.}},
\bauthor{\bsnm{{Schou}}, \binits{J.}},
\bauthor{\bsnm{{Thompson}}, \binits{M.J.}},
\bauthor{\bsnm{{Tomczyk}}, \binits{S.}},
\bauthor{\bsnm{{Toner}}, \binits{C.G.}},
\bauthor{\bsnm{{Toussaint}}, \binits{R.}},
\bauthor{\bsnm{{Williams}}, \binits{W.E.}}:
\byear{1996},
\batitle{{The Solar Acoustic Spectrum and Eigenmode Parameters}}.
\bjtitle{Science}
\bvolume{272},
\bfpage{1292}.
\doiurl{10.1126/science.272.5266.1292}.
\adsurl{https://ui.adsabs.harvard.edu/abs/1996Sci...272.1292H}.
\end{barticle}
\endbibitem

\bibitem[\protect\citeauthoryear{{Jain} \textit{et~al.}}{2012}]{jain2012}
\begin{barticle}
\bauthor{\bsnm{{Jain}}, \binits{K.}},
\bauthor{\bsnm{{Komm}}, \binits{R.W.}},
\bauthor{\bsnm{{Gonz{\'a}lez Hern{\'a}ndez}}, \binits{I.}},
\bauthor{\bsnm{{Tripathy}}, \binits{S.C.}},
\bauthor{\bsnm{{Hill}}, \binits{F.}}:
\byear{2012},
\batitle{{Subsurface Flows in and Around Active Regions with Rotating and
  Non-rotating Sunspots}}.
\bjtitle{\solphys}
\bvolume{279}(\bissue{2}),
\bfpage{349}.
\doiurl{10.1007/s11207-012-0039-0}.
\adsurl{https://ui.adsabs.harvard.edu/abs/2012SoPh..279..349J}.
\end{barticle}
\endbibitem

\bibitem[\protect\citeauthoryear{{Komm}}{2021}]{komm2021}
\begin{barticle}
\bauthor{\bsnm{{Komm}}, \binits{R.}}:
\byear{2021},
\batitle{{Subsurface Horizontal Flows During Solar Cycles 24 and 25 with
  Large-Tile Ring-Diagram Analysis}}.
\bjtitle{\solphys}
\bvolume{296}(\bissue{12}),
\bfpage{174}.
\doiurl{10.1007/s11207-021-01923-0}.
\adsurl{https://ui.adsabs.harvard.edu/abs/2021SoPh..296..174K}.
\end{barticle}
\endbibitem

\bibitem[\protect\citeauthoryear{{Lekshmi}, {Nandy}, and
  {Antia}}{2018}]{lekshmi2018}
\begin{barticle}
\bauthor{\bsnm{{Lekshmi}}, \binits{B.}},
\bauthor{\bsnm{{Nandy}}, \binits{D.}},
\bauthor{\bsnm{{Antia}}, \binits{H.M.}}:
\byear{2018},
\batitle{{Asymmetry in Solar Torsional Oscillation and the Sunspot Cycle}}.
\bjtitle{\apj}
\bvolume{861}(\bissue{2}),
\bfpage{121}.
\doiurl{10.3847/1538-4357/aacbd5}.
\adsurl{https://ui.adsabs.harvard.edu/abs/2018ApJ...861..121L}.
\end{barticle}
\endbibitem

\bibitem[\protect\citeauthoryear{{Scherrer} \textit{et~al.}}{1995}]{mdi}
\begin{barticle}
\bauthor{\bsnm{{Scherrer}}, \binits{P.H.}},
\bauthor{\bsnm{{Bogart}}, \binits{R.S.}},
\bauthor{\bsnm{{Bush}}, \binits{R.I.}},
\bauthor{\bsnm{{Hoeksema}}, \binits{J.T.}},
\bauthor{\bsnm{{Kosovichev}}, \binits{A.G.}},
\bauthor{\bsnm{{Schou}}, \binits{J.}},
\bauthor{\bsnm{{Rosenberg}}, \binits{W.}},
\bauthor{\bsnm{{Springer}}, \binits{L.}},
\bauthor{\bsnm{{Tarbell}}, \binits{T.D.}},
\bauthor{\bsnm{{Title}}, \binits{A.}},
\bauthor{\bsnm{{Wolfson}}, \binits{C.J.}},
\bauthor{\bsnm{{Zayer}}, \binits{I.}},
\bauthor{\bsnm{{MDI Engineering Team}}}:
\byear{1995},
\batitle{{The Solar Oscillations Investigation - Michelson Doppler Imager}}.
\bjtitle{\solphys}
\bvolume{162},
\bfpage{129}.
\doiurl{10.1007/BF00733429}.
\adsurl{https://ui.adsabs.harvard.edu/abs/1995SoPh..162..129S}.
\end{barticle}
\endbibitem

\bibitem[\protect\citeauthoryear{{Scherrer} \textit{et~al.}}{2012}]{HMI}
\begin{barticle}
\bauthor{\bsnm{{Scherrer}}, \binits{P.H.}},
\bauthor{\bsnm{{Schou}}, \binits{J.}},
\bauthor{\bsnm{{Bush}}, \binits{R.I.}},
\bauthor{\bsnm{{Kosovichev}}, \binits{A.G.}},
\bauthor{\bsnm{{Bogart}}, \binits{R.S.}},
\bauthor{\bsnm{{Hoeksema}}, \binits{J.T.}},
\bauthor{\bsnm{{Liu}}, \binits{Y.}},
\bauthor{\bsnm{{Duvall}}, \binits{T.L.}},
\bauthor{\bsnm{{Zhao}}, \binits{J.}},
\bauthor{\bsnm{{Title}}, \binits{A.M.}},
\bauthor{\bsnm{{Schrijver}}, \binits{C.J.}},
\bauthor{\bsnm{{Tarbell}}, \binits{T.D.}},
\bauthor{\bsnm{{Tomczyk}}, \binits{S.}}:
\byear{2012},
\batitle{{The Helioseismic and Magnetic Imager (HMI) Investigation for the
  Solar Dynamics Observatory (SDO)}}.
\bjtitle{\solphys}
\bvolume{275},
\bfpage{207}.
\doiurl{10.1007/s11207-011-9834-2}.
\adsurl{https://ui.adsabs.harvard.edu/abs/2012SoPh..275..207S}.
\end{barticle}
\endbibitem

\bibitem[\protect\citeauthoryear{{Schou} \textit{et~al.}}{2012}]{hmical}
\begin{barticle}
\bauthor{\bsnm{{Schou}}, \binits{J.}},
\bauthor{\bsnm{{Scherrer}}, \binits{P.H.}},
\bauthor{\bsnm{{Bush}}, \binits{R.I.}},
\bauthor{\bsnm{{Wachter}}, \binits{R.}},
\bauthor{\bsnm{{Couvidat}}, \binits{S.}},
\bauthor{\bsnm{{Rabello-Soares}}, \binits{M.C.}},
\bauthor{\bsnm{{Bogart}}, \binits{R.S.}},
\bauthor{\bsnm{{Hoeksema}}, \binits{J.T.}},
\bauthor{\bsnm{{Liu}}, \binits{Y.}},
\bauthor{\bsnm{{Duvall}}, \binits{T.L.}},
\bauthor{\bsnm{{Akin}}, \binits{D.J.}},
\bauthor{\bsnm{{Allard}}, \binits{B.A.}},
\bauthor{\bsnm{{Miles}}, \binits{J.W.}},
\bauthor{\bsnm{{Rairden}}, \binits{R.}},
\bauthor{\bsnm{{Shine}}, \binits{R.A.}},
\bauthor{\bsnm{{Tarbell}}, \binits{T.D.}},
\bauthor{\bsnm{{Title}}, \binits{A.M.}},
\bauthor{\bsnm{{Wolfson}}, \binits{C.J.}},
\bauthor{\bsnm{{Elmore}}, \binits{D.F.}},
\bauthor{\bsnm{{Norton}}, \binits{A.A.}},
\bauthor{\bsnm{{Tomczyk}}, \binits{S.}}:
\byear{2012},
\batitle{{Design and Ground Calibration of the Helioseismic and Magnetic Imager
  (HMI) Instrument on the Solar Dynamics Observatory (SDO)}}.
\bjtitle{\solphys}
\bvolume{275}(\bissue{1-2}),
\bfpage{229}.
\doiurl{10.1007/s11207-011-9842-2}.
\adsurl{https://ui.adsabs.harvard.edu/abs/2012SoPh..275..229S}.
\end{barticle}
\endbibitem

\bibitem[\protect\citeauthoryear{{Sekii}}{1997}]{sekii1997}
\begin{bchapter}
\bauthor{\bsnm{{Sekii}}, \binits{T.}}:
\byear{1997},
\bctitle{{Internal Solar rotation}}.
In: \beditor{\bsnm{{Provost}}, \binits{J.}},
\beditor{\bsnm{{Schmider}}, \binits{F.-X.}} (eds.)
\bbtitle{Sounding Solar and Stellar Interiors}
\bseriesno{181},
\bfpage{ISBN0792348389}.
\adsurl{https://ui.adsabs.harvard.edu/abs/1997IAUS..181..189S}.
\end{bchapter}
\endbibitem

\bibitem[\protect\citeauthoryear{{Snodgrass}}{1984}]{snodgrass}
\begin{barticle}
\bauthor{\bsnm{{Snodgrass}}, \binits{H.B.}}:
\byear{1984},
\batitle{{Separation of large-scale photospheric Doppler patterns}}.
\bjtitle{\solphys}
\bvolume{94}(\bissue{1}),
\bfpage{13}.
\doiurl{10.1007/BF00154804}.
\adsurl{https://ui.adsabs.harvard.edu/abs/1984SoPh...94...13S}.
\end{barticle}
\endbibitem

\end{thebibliography}

\end{article} 

\end{document}